\documentclass[11pt]{article}
\usepackage[T1]{fontenc}
\usepackage[latin9]{inputenc}
\usepackage{color, colortbl}
\usepackage[table]{xcolor}
\usepackage{bm}
\usepackage{amsmath}
\usepackage{amssymb}
\usepackage{graphicx}
\usepackage{tikz}
\usepackage{booktabs}
\usepackage{subcaption}
\usetikzlibrary{fit,positioning}
\usepackage{tikz-cd}
\usepackage{caption}
\usepackage{hhline}
\usepackage{mathtools}

\usepackage[authoryear, round]{natbib}
\usepackage[unicode=true,
 bookmarks=false,
 breaklinks=false,pdfborder={0 0 1},backref=section,colorlinks=false]
 {hyperref}

\makeatletter
\allowdisplaybreaks
\providecommand{\tabularnewline}{\\}

\@ifundefined{date}{}{\date{}}

\usepackage{latexsym}
\usepackage{multirow}\usepackage{bm}
\usepackage{url}

\usepackage{enumerate}
\usepackage{authblk}

\usepackage{dcolumn}
\newcolumntype{.}{D{.}{.}{-1}}
\newcolumntype{d}[1]{D{.}{.}{#1}}
\usepackage{theorem}
\theoremstyle{definition}
\newtheorem{assumption}{Assumption}\newtheorem{theorem}{Theorem}\newtheorem{lemma}{Lemma}\newtheorem{condition}{Condition}

\def\independenT#1#2{\mathrel{\rlap{$#1#2$}\mkern2mu{#1#2}}}

\newcommand{\indep}{\protect\mathpalette{\protect\independenT}{\perp}}
\def\independenT#1#2{\mathrel{\rlap{$#1#2$}\mkern2mu{#1#2}}}
\definecolor{Gray}{gray}{0.9}

\usepackage{rotating}

\usepackage{arydshln}

\usepackage[compact]{titlesec}

\allowdisplaybreaks

\newcommand{\spacingset}[1]{\renewcommand{\baselinestretch}%
{#1}\small\normalsize}

\newcommand{\pr}{P}

\newcommand{\R}{\mathbb{R}}

\newcommand{\T}{\mathrm{\scriptscriptstyle T}}
\newcommand{\logit}{\text{logit}}

\newcommand{\J}{\mathcal{J}}
\newcommand{\N}{\mathcal{N}}

\newcommand{\ipsw}{\mathrm{IPSW}}
\newcommand{\aipsw}{\mathrm{AIPSW}}

\newcommand{\bbeta}{\bm{\beta}}
\newcommand{\btau}{\bm{\tau}}

\newcommand{\htau}{\hat{\tau}}
\newcommand{\blam}{\bm{\lambda}}
\newcommand{\bg}{\bm{g}}
\newcommand{\bEta}{\bm{\eta}}
\newcommand{\bpsi}{\bm{\psi}}
\newcommand{\btheta}{\bm{\theta}}

\newcommand{\tildelta}{\widetilde{\delta}}

\newcommand{\M}{\mathcal{M}}

\def\bSig\mathbf{\Sigma}

\makeatother

\begin{document}
\title{\Large \textbf{Improving trial generalizability using observational studies}}
\author
{\small \textbf{Dasom Lee$^1$, Shu Yang$^{1,*}$, Lin Dong$^1$, Xiaofei Wang$^{2}$, Donglin Zeng$^{3}$, and Jianwen Cai$^{3}$}\\
$^{1}$Department of Statistics, North Carolina State University, Raleigh, NC, U.S.A. \\
\vspace{1ex} $^{2}$Department of Biostatistics and Bioinformatics, Duke University, Durham, NC, U.S.A. \\
\vspace{1ex} $^{3}$Department of Biostatistics, University of North Carolina at Chapel Hill, Chapel Hill, 
NC, U.S.A.\\
\vspace{1ex} $^*$\textit{email:} syang24@ncsu.edu
}


\maketitle

\spacingset{1.5} 
\begin{abstract}
Complementary features of randomized controlled trials (RCTs) and observational studies (OSs) can be used jointly to estimate the average treatment effect of a target population. We propose a calibration weighting estimator that enforces the covariate balance between the RCT and OS, therefore improving the trial-based estimator's generalizability. 
Exploiting semiparametric efficiency theory, we propose
a doubly robust augmented calibration weighting estimator that achieves the efficiency bound derived under the identification assumptions. A nonparametric sieve method is provided as an alternative to the parametric approach, which enables the robust approximation of the nuisance functions
and data-adaptive selection of outcome predictors for calibration. We establish
asymptotic results and confirm the finite sample performances of
the proposed estimators by simulation experiments and an application
on the estimation of the treatment effect of adjuvant chemotherapy for early-stage non-small cell lung patients after surgery.

\end{abstract}
\newpage{}

\section{Introduction}
\label{sec:Introduction} 
Randomized controlled trials (RCTs) are the
gold standard to evaluate treatment effects. However, due
to restrictive inclusion and exclusion criteria, 
the trial sample is narrowly defined and can be systematically different
from the real-world patient population to which the new treatment
is supposed to be given. Therefore, the findings from RCTs often lack
external validity \citep{rothwell2005external}. 
On the other hand, observational studies (OSs) often include large samples that are representative
of real-world patient populations; however, there are concerns about
whether or not confounding has been addressed adequately in the analyses
of OSs. In cancer research, there is an in-depth
discussion on the strengths and limitations of utilizing data from
RCT and OSs for comparative effectiveness analyses 
\citep{korn2012methodology}.

The problems of extending findings from RCT to a target population
has been termed as \textit{generalizability} \citep[e.g.,][]{cole2010generalizing, tipton2013improving,  dahabreh2019generalizing}  
and  \textit{transportability}
\citep[e.g.,][]{pearl2011transportability,rudolph2017robust, westreich2017transportability}.
Most existing methods rely
on direct modeling of the sampling score, 
the sampling analog of the propensity score.   
The subsequent
sampling score adjustments include inverse probability of sampling
weighting (IPSW; \citealp{cole2010generalizing,buchanan2018generalizing}),
stratification \citep{tipton2013improving}, and
augmented IPSW \citep[AIPSW;][]{dahabreh2019generalizing}. 
Most sampling score adjustment approaches require the sampling score model to be correctly specified. Moreover, weighting estimators  are unstable if the sampling score
is too extreme.

We consider combining an RCT sample and an OS
sample to estimate the average treatment effect (ATE) of a target
population, where the RCT sample is subject to selection bias and
the OS sample is representative of the target population
with a known sampling mechanism. 
In contrast to the dominant approaches that focus on predicting sample selection probabilities, we estimate the sampling score weights
directly by calibrating covariates balance between the RCT sample and
the design-weighted OS sample to address the
selection bias of the RCT sample.
Calibration weighting (CW) is widely used to integrate auxiliary information
in survey sampling  \citep{wu2001model} 
and causal inference 
\citep{qin2007empirical, hainmueller2012entropy}. 
\citet{hartman2015sample}
implemented CW to estimate the population ATEs by combining RCTs
with OSs.

The efficiency of the CW estimator can be further improved.
We derive the semiparametric efficiency
bound for the ATE under the identification assumptions, which provides
the benchmark for estimation efficiency. We propose the augmented
CW (ACW) estimator that is doubly robust and also achieves the semiparametric
efficiency bound when both nuisance models are correctly specified.
However, the parametric approach is prone to model misspecification,
especially when there is complex confounding. To cope with model misspecification, we adopt a method
of sieves \citep{chen2007large}, which allows flexible
data-adaptive estimation of the nuisance functions while the ACW estimator retains
the usual root-$n$ consistency under regularity conditions.
In comparison with other nonparametric and machine
learning methods, the proposed ACW estimator with the sieve approximation is attractive: 1) 
unlike black-box machine learning methods, calibration weighting is
straightforward and transparent; and 2)  our framework allows for selecting important sieve basis terms that are related to the outcome to calibrate and enforcing the balance on these covariates for efficient estimation.

In the presence of many covariates, variable or sieve basis selection for
calibration becomes necessary. {We classify covariates into three types: the covariates that are associated with both trial participation and outcome as confounders, that affect outcome only through trial participation as instrumental variables (IVs), and that are predictive of the outcome as precision variables or outcome predictors. 
In other causal inference contexts, studies have shown that in addition to the confounding variables, including outcome predictors in the propensity score may improve efficiency, whereas including IVs may decrease efficiency \citep[e.g., ][]{tang2020outcome}.  
Despite the importance of proper basis selection for the efficient causal estimator, the current literature lacks a principled approach to guide basis selection for covariate balancing. 
Capitalizing on an explicit connection between
calibration weighting and estimating equations under parametric models, 
we propose a penalized estimating equation approach for
variable selection with an emphasis on outcome predictors.

\section{Basic Setup} \label{sec:basic setup}
\subsection{Notation: causal effect and two data sources}
\label{sec:notation}
Let $X$ be the $p$-dimensional vector of covariates, $A$
be the binary treatment $\{0,1\}$,
and $Y$ be the outcome of interest.  
We use the potential outcomes framework 
to formulate the causal problem. 
We assume that each subject 
has a potential
outcome $Y(a), a \in \{0,1\}$, representing the outcome had the subject been given the treatment $a$. The conditional average
treatment effect (CATE) is $\tau(X)=E\{Y(1)-Y(0)\mid X\}$.
We are interested in estimating the population ATE $\tau_{0}=E\{\tau(X)\}$,
where the expectation is taken with respect to the distribution of the target population. 

Let $\delta=1$
denote RCT participation, and let $\tildelta=1$ denote the OS participation. 
Also, define the sampling score as $\pi_{\delta}(X)=\pr(\delta=1\mid X)$, the design weight for the OS sample as $d=1/P(\tildelta=1\vert X)$,
and the conditional outcome mean function as $\mu_{a,\delta}(X)=E(Y\mid X,A=a,\delta)$
for $a,\delta\in\{0,1\}$.

To generalize findings to the future
patient population, we consider a super-population framework that
describes the distribution of all patients with a certain disease
to whom the new treatment is intended to be given. The data structure is demonstrated
in Figure \ref{fig:RCTRWEdemo}. The RCT is a sample
from the target population with an unknown sampling mechanism, and
the OS sample is a random sample from
the target population with a known sampling mechanism. 
Therefore, our problem is in line with that of generalizability, extending the ATE result from the trial to its larger population \citep{dahabreh2019generalizing}. A closely related problem is transportability which tries to extend the trial results to an external population \citep{westreich2019target}; 
e.g., when one wants to transport an RCT conducted in one country to a population
in another country \citep{pearl2015findings}.
Subtle differences exist in the two problems in terms of estimands and identification assumptions; 
see Section \ref{s:gen_trans} of the supplementary material for details.
In general, there are nested and non-nested study designs in the problem of generalization \citep{dahabreh2019study}.
For nested designs, the RCT sample is a subsample from the OS
sample. Examples include pragmatic trial studies embedded in health
care systems or comprehensive cohort studies, where all trial-eligible
participants constitute the OS sample, and participants
who agree to be randomized constitute the RCT sample. For non-nested
designs, the RCT sample and the OS sample are separate.
Our motivating application falls in the latter category, where we link an
existing RCT to a large cancer register database. 

We also assume that the RCT and OSs are independent.
This assumption holds naturally if the two separate studies are conducted
independently by different researchers; it is also plausible in our
motivating example where the patients for the two studies were accrued
in two separate time periods (see Section \ref{sec:Real-data-application}).

\begin{figure}[h]
\begin{centering}

\begin{tikzpicture}[every node/.style=scale=1.1, node distance = 3cm, auto]
\tikzstyle{block} = [rectangle, draw, text width=18em, text centered, rounded corners, minimum height=4em]
\tikzstyle{every node} = [font = \footnotesize]
\node [block] (A) {	\begin{tabular}{c} 	
    \textbf{Target future patient population} \\
    	(Superpopulation)\\
	$\left\{X_i,Y_i(0),Y_i(1)\right\}_{i=1}^\infty$
\end{tabular}
};
\node at (-4.5,-2) [block] (B) {	\begin{tabular}{c} 	
    \textbf{Finite RCT population}
\end{tabular}
};
\node at (4.5,-2) [block] (C) {	\begin{tabular}{c} 	
    \textbf{Finite OS population}
\end{tabular}
};
\node [block, below of=B] (D) {
	\begin{tabular}{l} 	
	$\left\{ X_i,Y_i(0),Y_i(1), \delta_i = 1,\widetilde{\delta}_i = 0 \right\}_{i=1}^n$
	\end{tabular}
};
\node  [block, below of = C] (E) {\begin{tabular}{l} 
	$\left\{X_i,Y_i(0),Y_i(1),  \delta_i =0,\widetilde{\delta}_i = 1\right\}_{i=n+1}^{n+m}$
\end{tabular}
};
\node at (-4.5,-3.5)(s-RCT){RCT Sampling: $\delta_i \sim$ Unknown};
\node at (4.5,-3.5)(s-RWE){OS Sampling: $\widetilde{\delta}_i\sim $ Known design};
\node at (-4.5,-6.5)(t-RCT) {RCT Treatment: $A_i \sim$ Randomization};
\node at (4.5,-6.5)(t-RWE) {OS Treatment: $A_i \sim$ Unknown};
\node [block, below of=D] (F) {
	\begin{tabular}{c}
	\textbf{Observed RCT Sample}\\
	$\left\{X_i,A_i,Y_i, \delta_i = 1,\widetilde{\delta}_i = 0\right\}_{i=1}^n$
	\end{tabular}
};
\node [block, below of=E] (G) {
\begin{tabular}{c}
 \textbf{Observed OS Sample}\\
$\left\{X_i,(A_i,Y_i),  \delta_i = 0,\widetilde{\delta}_i=1\right\}_{i=n+1}^{n+m}$
\end{tabular}
};
\draw [->,shorten >=4pt] 

(A) edge (B) 
(A) edge (C)
(B) edge (s-RCT)
(s-RCT) edge (D)
(C) edge (s-RWE)
(s-RWE) edge (E)
(D) edge (t-RCT)
(E) edge (t-RWE)
(t-RCT) edge (F)
(t-RWE) edge (G); 
\end{tikzpicture}
\par\end{centering}
\caption{\label{fig:RCTRWEdemo}Demonstration of the sampling and treatment
assignment regimes for the RCT and OS samples within the
target population.}
\end{figure}
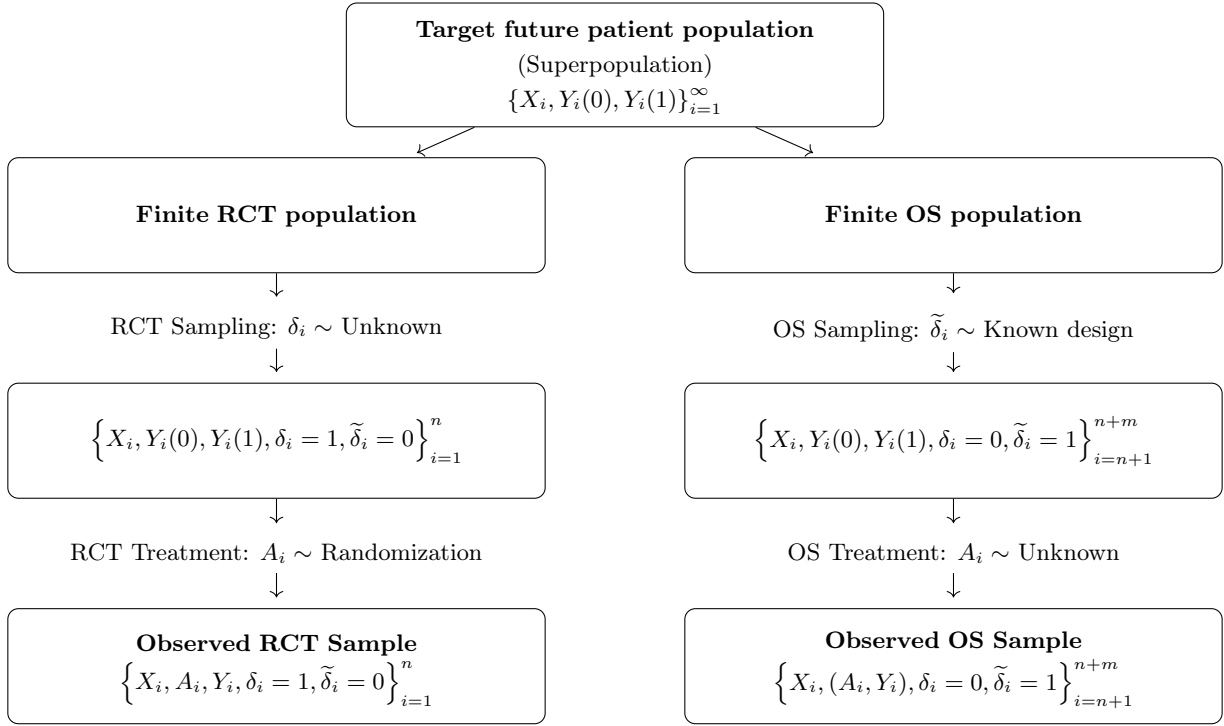

\subsection{Identification assumptions}
\label{sec:identification}

To identify
the ATE, we make the following assumptions.
\begin{assumption}[Consistency]\label{a:consist}
$Y=AY(1)+(1-A)Y(0)$.
\end{assumption}

\begin{assumption}[Ignorability and positivity of treatment assignment]\label{a:ign}  \textcolor{white}{s}\\
(i) $\{Y(0),Y(1)\} \indep A\mid(X,\delta=1)$; and (ii) $0<P(A=1\mid X,\delta=1)<1$ with probability $1$.
\end{assumption}

\begin{assumption}[Generalizability of the CATE and positivity of trial participation]\label{a:pos}
 \textcolor{white}{s}\\ (i) $E\{Y(1) - Y(0)\mid X, \delta = 1\}=\tau(X)$; 
and (ii)
$\pi_{\delta}(X)>0$  with probability $1$.
\end{assumption}

Assumption \ref{a:consist}
implies that trial encouragement effects are absent (\citealp{dahabreh2019extending}). 
Assumption \ref{a:ign}  holds for the RCT by default. 
Assumption \ref{a:pos} (i) is similar to the generalizability in effect measure condition \citep[][Supporting Information]{dahabreh2019generalizing}. Even though this assumption is formally weaker than the mean exchangeability over trial participation \citep{dahabreh2019generalizing},
i.e., $E\{Y(a)\mid X,\delta=1\}=E\{Y(a)\mid X\}$ for $a=0,1$, and the ignorability assumption on trial participation \citep{stuart2011use}, i.e.,  $\{Y(0),Y(1)\}\indep\delta\mid X$, it suffices to identify the ATE.
Under Assumptions \ref{a:consist}--\ref{a:pos}, the ATE is identified by 
$\tau_{0}=E\left[\tildelta d\{\mu_{1,1}(X)-\mu_{0,1}(X)\}\right].$

Although being essential, Assumption \ref{a:pos} (i)  is not verifiable based on the observed data but relies on subject matter experts to assess its plausibility. It is plausible if $X$
captures all variables that are related to the trial participation and outcome 
\citep{buchanan2018generalizing}. 
Assumption \ref{a:pos} (ii) 
requires the absence
of patient characteristics that prohibit participation.
When Assumption \ref{a:pos} (ii) is violated, generalization
can only be made to a restricted population without extrapolation
\citep{yang2018trimming}.

\subsection{Existing estimation methods}
Because the RCT assigns treatments randomly to the participants,
$\tau(X)$ is identifiable and can be estimated by standard estimators
solely from the RCT. However, 
$f(X\mid\delta=1)$ is different from 
$f(X)$ in general; therefore, $E\{\tau(X)\mid\delta=1\}$ is different
from $\tau_{0}$, and the ATE estimator using trial data only is biased
of $\tau_{0}$ generally. A widely-used approach is the IPSW estimator that predicts the sampling
score $\pi_{\delta}(X)$ and uses the inverse of the estimated sampling
score to account for the shift of the covariate distribution from
the RCT sample to the target population. Specifically, most of the
empirical literature assumes that $\pi_{\delta}(X)$ follows a logistic
regression model $\pi_{\delta}(X;\bEta)$ and can be estimated by
$\pi_{\delta}(X;\widehat{\bEta})$. 
The AIPSW estimator has also been proposed to improve it by employing both the sampling
score and outcome regression. The forms of the (A)IPSW estimators are provided in Section \ref{s:ipsw_est} of the supplementary material, along with another identification formula based on the IPSW estimator.

\section{Calibration Weighting Estimator}
\label{sec: CWmethod}
We propose to use calibration originated in survey sampling to eliminate
the selection bias in the trial-based ATE estimator. The calibration
weighting approach is similar to the idea of entropy balancing weights
introduced by \citet{hainmueller2012entropy}. We calibrate subjects
in the RCT sample so that after calibration, the covariate distribution
of the RCT sample empirically matches the target population. Our insight
is that for any vector-valued function $\bg(X)$, the following equations hold:
\[
E\left[\frac{\{\pi_{\delta}(X)\}^{-1}\delta}{E[\{\pi_{\delta}(X)\}^{-1}\delta]}\bg(X)\right]=E\left\{ \tildelta d\bg(X)\right\} =E\{\bg(X)\}.
\]
Here, $\bg(X)$ contains the covariate functions to be calibrated,
which could be moment functions of the original covariate $X$ or
any sensible transformations of $X$.
To this end, we assign a weight $q_{i}$ to each subject $i$ in the
RCT sample so that 
\begin{equation}
\sum_{i=1}^{N}\delta_{i}q_{i}\bg(X_{i})=\widetilde{\bg},\label{eq:calibration constraints}
\end{equation}
where $\widetilde{\bg}=\sum_{i=1}^{N}\tildelta_{i}d_{i}\bg(X_{i})/\sum_{i=1}^{N}\tildelta_{i}d_{i}$
is a design-weighted estimate of $E\{\bg(X)\}$ from the OS
sample, and $N$ is the target population size, not necessarily known. Constraint \eqref{eq:calibration constraints} is referred
to as the balancing constraint, and weights $\mathcal{Q}=\{q_{i}:\delta_{i}=1\}$
are the calibration weights. The balancing constraint calibrates the
RCT covariate distribution  to the target population
in terms of $\bg(X)$. The choice of $\bg(X)$ is critical for both
bias and variance considerations, which we discuss in Section
\ref{sec:Choice-of-g}.

We estimate $\mathcal{Q}$ by solving the optimization problem:
\begin{equation}
\underset{\mathcal{Q}}{\text{min}}\sum_{i=1}^{n}q_{i}\log q_{i},\label{eq:optQ}
\end{equation}
subject to $q_{i}\ge0$ $\forall i$; $\sum_{i=1}^{n}q_{i}=1$,
and the constraint (\ref{eq:calibration constraints}), where $n$ is the RCT sample size.
The objective function in \eqref{eq:optQ} is the negative entropy of the calibration
weights; thus, minimizing these criteria ensures that the empirical
distribution of calibration weights is not too far away from the
uniform, such that it minimizes the variability due to heterogeneous
weights \citep{owen2001empirical}. 
The optimization problem can be solved using convex optimization
with the Lagrange multiplier. By introducing the Lagrange multiplier $\blam$,
the objective function becomes 
\begin{equation}
L(\blam,\mathcal{Q})=\sum_{i=1}^{n}q_{i}\log q_{i}-\blam^{\top}\left\{ \sum_{i=1}^{n}q_{i}\bg(X_{i})-\widetilde{\bg}\right\} .\label{eq:lagrange_obj}
\end{equation}
Minimizing \eqref{eq:lagrange_obj} leads to
$
\widehat{q}_{i}=q(X_{i};\widehat{\blam})={\exp\{ \widehat{\blam}^{\top}\bg(X_{i})\} }/{\sum_{i=1}^{n}\exp\{ \widehat{\blam}^{\top}\bg(X_{i})\} },
$
and $\widehat{\blam}$ solves
\begin{equation}
U(\blam)=\sum_{i=1}^{n}\exp\left\{ \blam^{\top}\bg(X_{i})\right\} \left\{ \bg(X_{i})-\widetilde{\bg}\right\} =0,\label{eq:lam}
\end{equation}
which is the dual problem to the optimization problem (\ref{eq:optQ}).

Let $\pi_{Ai}=P(A_{i}=1\vert X_{i},\delta_{i}=1)$ be the treatment
propensity score for subject $i$. For RCTs, it is common that the
propensity score $\pi_{Ai}$ is known.
 The CW estimator becomes
\begin{equation}
\htau^{\mathrm{CW}}=\sum_{i=1}^{n}\widehat{q}_{i}\left\{ \frac{A_{i}Y_{i}}{\pi_{Ai}}-\frac{(1-A_{i})Y_{i}}{1-\pi_{Ai}}\right\}. \label{eq:cbss0}
\end{equation}

To investigate the properties of the CW estimator, we impose
the regularity conditions on the sampling designs for 
the RCT the OS samples.

\begin{assumption}\label{assump:design_wt}
Let $\boldsymbol{\mu}_{\bg0}=E\{\bg(X)\}$. The design weighted estimator
$\widehat{\boldsymbol{\mu}}_{\bg}=N^{-1}\sum_{i=1}^{N}\tildelta_{i}d_{i}\bg(X_{i})$
satisfies $V(\widehat{\boldsymbol{\mu}}_{\bg})=O(m^{-1})$, and $\{V(\widehat{\boldsymbol{\mu}}_{\bg})\}^{-1/2}(\widehat{\boldsymbol{\mu}}_{\bg}-\boldsymbol{\mu}_{\bg_{0}})\rightarrow\mathcal{N}(0,1)$
in distribution, as $m\rightarrow\infty$, where $m$ is the OS sample size.
\end{assumption}

\begin{assumption}[Linearity of the CATE]
\label{assump: outcome linear}
$\tau(X)=\boldsymbol{\gamma}_{0}^{\top}\bg(X)$.
\end{assumption}

\begin{assumption}[Loglinear sampling score]\label{asump:logit linear}
The sampling score of RCT participation follows a loglinear model,
i.e., $\pi_{\delta}(X)=\exp\{\bEta_{0}^{\top}\bg(X)\}$ for some $\bEta_{0}$.
\end{assumption}
Based on the above assumptions, we establish the double robustness
property of the CW estimator in the following theorem and relegate
all proofs to Section \ref{s:proof} of the supplementary material. The proof is similar to
the one in \citet{zhao2017entropy}.
\begin{theorem}[Double robustness of the CW estimators]\label{th:consistency1}
Under Assumptions \ref{a:consist}--\ref{assump:design_wt}, if either
Assumption \ref{assump: outcome linear} or \ref{asump:logit linear} holds, 
$\hat{\tau}^{\mathrm{CW}}$ in (\ref{eq:cbss0})
is consistent for $\tau_{0}$.
\end{theorem}

In the estimation of calibration weights, we only require
specifying $\bg(X)$. Thus, calibration weighting evades explicitly
modeling either the sampling score model or the outcome mean models.
Under Assumption \ref{asump:logit linear}, we show that there is
a direct correspondence between calibration weight $q(X_{i};\widehat{\blam})$
and the estimated sampling score $\pi_{\delta}(X_{i};\widehat{\bEta})$,
i.e. $q(X_{i};\widehat{\blam})=\left\{ N\pi_{\delta}(X_{i};\widehat{\bEta})\right\} ^{-1}+o_{p}(N^{-1})$.
That is, calibration weights from the objective function \eqref{eq:optQ} have the same functional form as inverse probability of sampling score weights under Assumption \ref{asump:logit linear} asymptotically. 

Other objective functions, such as $\sum_{i=1}^n\left(q_{i} - 1 \right) \log \left(q_{i} - 1 \right)$, $\sum_{i=1}^n \{q_{i}\log q_{i} + \left(1 - q_{i} \right) \log \left(1 - q_{i} \right)\}$ \citep{zhao2019covariate, josey2020framework} or $\sum_{i=1}^n\left (q_i - {n}^{-1} \right)^2$ \citep{chattopadhyay2020balancing}, can also be used. When the sampling score follows a logistic regression model,
the objective function $\sum_{i=1}^n\left(q_{i} - 1 \right) \log \left(q_{i} - 1 \right)$ results in weights that resemble the inverse of logistic sampling scores \citep{zhao2019covariate, josey2020framework}. However, if the fraction $n/N$ of the RCT sample in the target population
is small, the loglinear model in Assumption \ref{asump:logit linear}
is close to the logistic regression model;  our simulation studies show that the proposed CW estimator is not sensitive to the choice of the objective function for the optimization.

The entropy balancing has been studied in the indirect comparison literature \citep{signorovitch2010comparative, phillippo2018methods, petto2019alternative}. The goal is to adjust for the imbalance between two separate randomized trials with common comparative arms, similar to the transportability problem. On the other hand, the proposed CW estimator is motivated by generalizing findings from RCT. 
Importantly, building on the CW estimator, we propose an improved estimator capitalizing semiparametric efficiency theory in the next section and a data-adaptive selection of outcome predictors for calibration, which is absent in the literature.
The proposed framework can incorporate
nonparametric sieve approximation of the outcome mean function and
sampling score while providing valid inferences.

\section{Semiparametric Efficient Estimator}
\label{sec:Efficiency-boosting-through}
\subsection{Augmented calibration weighting estimator}
\label{sec:ACW}

The following theorem gives the semiparametric efficiency bound for
$\tau_{0}$ in our data integration setting. Let ${\Delta}_{a} = Y -  \mu_{a,1}(X;{\bbeta}_{a})$.

\begin{theorem}[Semiparametric efficiency bound]\label{th:semi_bd-1}
Under Assumptions \ref{a:consist} - \ref{assump:design_wt}, the
semiparametric efficiency score for $\tau_{0}$ is 
\begin{equation*}
    \phi(X,A,Y,\delta,\widetilde{\delta})=\frac{\delta}{\pi_{\delta}(X)}\left[\frac{A{\Delta}_{1} }{\pi_{A}}-\frac{(1-A){\Delta}_{0} }{1-\pi_{A}}\right] \\
    + \tildelta d\left\{ \tau(X)-\tau_{0}\right\}.
\end{equation*}
The semiparametric efficiency bound for $\tau_{0}$ is 
\begin{equation*}
    V_{\mathrm{eff}}=E\left[\frac{\delta}{\pi_{\delta}(X)^{2}}\left\{ \frac{V\left\{ Y(1)|X,\delta\right\} }{\pi_{A}}+\frac{V\left\{ Y(0)|X,\delta\right\} }{1-\pi_{A}}\right\}\right. \\+
    \left.\tildelta d^{2}\left\{ \tau(X)-\tau_{0}\right\} ^{2}\right].
\end{equation*}
\end{theorem}

The result in Theorem \ref{th:semi_bd-1} serves as a foundation to
derive efficient estimators combining two data sources. Under Assumption
\ref{a:ign}, $\tau(X)=\mu_{1,1}(X)-\mu_{0,1}(X)$.
The score $\phi(X,A,Y,\delta,\widetilde{\delta})$ has unknown nuisance
functions $\pi_{\delta}(X)$ and $\mu_{a,1}(X)$, $a=0,1$.
Therefore, to estimate $\tau_{0}$, we posit models for the nuisance
functions, denoted by $\pi_{\delta}(X;\bEta)$ and $\mu_{a,1}(X;\bbeta_{a})$.
For example, we assume $\pi_{\delta}(X)$ is a loglinear model as
in Assumption \ref{asump:logit linear}. By the correspondence between
the loglinear model and the calibration weighting algorithm, we can
estimate $\bEta_{0}$ following the optimization algorithm in \eqref{eq:optQ}.
We also posit models $\mu_{a,1}(X;\bbeta_{a}),\;a=0,1$. By
Assumption \ref{a:ign}, we are able to obtain a consistent estimator
$\widehat{\bbeta}_{a}$ based on the trial sample. Based on the semiparametric efficiency score, we propose a new estimator
for the ATE. As the outcome mean models in the semiparametric efficiency
score can be viewed as an augmentation to the CW estimator, we refer
to the proposed estimator as the augmented calibration weighting (ACW)
estimator. Let $\hat{\Delta}_{a,i} = Y_i -  \mu_{a,1}(X_{i};\widehat{\bbeta}_{a})$. The ACW estimator is 
\begin{equation}\label{eq:taudc-1}
  \htau^{\mathrm{ACW}}=\sum_{i=1}^{N}\delta_{i}\widehat{q}_{i}\left\{\frac{A_{i}\hat{\Delta}_{1,i} }{\pi_{Ai}}
-\frac{(1-A_{i})\hat{\Delta}_{0,i} }{1-\pi_{Ai}}\right\} + \\
\left( \sum_{i=1}^{N}\tildelta_{i}d_{i}\right) ^{-1}\sum_{i=1}^{N}\tildelta_{i}d_{i}\left\{ \mu_{1,1}(X_{i};\widehat{\bbeta}_{1})-\mu_{0,1}(X_{i};\widehat{\bbeta}_{0})\right\} .  
\end{equation}
We now show  that $\htau^{\mathrm{ACW}}$ achieves double robustness and local efficiency. For a vector $v$, we use
$\Vert v\Vert_{2}=(v^{\top}v)^{1/2}$ to denote its Euclidean norm.
For a function $f(V)$, where $V$ is a generic random variable, we
define its $L_{2}$-norm as $\Vert f(V)\Vert=\left\{ \int f(v)^{2}dP(v)\right\} ^{1/2}$.

\begin{theorem}[Double robustness and local efficiency]
\label{th:consistency2}
Under Assumptions \ref{a:consist}--\ref{assump:design_wt}, if either
Assumptions \ref{assump: outcome linear} or \ref{asump:logit linear} 
holds,
$\htau^{\mathrm{ACW}}$
is consistent for $\tau_{0}$. When both assumptions hold, $N^{1/2}(\htau^{\mathrm{ACW}}-\tau_{0})\rightarrow\mathcal{N}(0,V_{\mathrm{eff}})$
in distribution, as $n\rightarrow\infty$, where $V_{\mathrm{eff}}$
is defined in Theorem \ref{th:semi_bd-1}, i.e., $\htau^{\mathrm{ACW}}$
is locally efficient.
\end{theorem}

By the empirical processes theory, the effect of nuisance parameter
estimation in $\htau^{\mathrm{ACW}}-\tau_{0}$ is bounded by $\lVert\pi_{\delta}(X;\widehat{\bEta})-\pi_{\delta}(X)\rVert\sum_{a=0}^{1}\rVert\mu_{a,1}(X;\widehat{\bbeta}_{a})-\mu_{a,1}(X)\rVert$;
see Section \ref{subsec:localEff} of the supplementary material for details. 
If this bound is  of rate $o_{p}(n^{-1/2})$, it
is asymptotically negligible. Thus, $\htau^{\mathrm{ACW}}$ is
semiparametric efficient. 
In general, there exist
different combinations of convergence rates of $\pi_{\delta}(X;\widehat{\bEta})$
and $\mu_{a,1}(X;\widehat{\bbeta}_{a})\;(a=0,1)$ that result
in a negligible error bound accommodating different smoothness conditions
of the underlying nuisance functions. The following theorem formalizes
the above statement.

\begin{theorem}\label{th:semi_nuisance}
Suppose Assumptions \ref{a:consist}--\ref{assump:design_wt}
hold. Let $\pi_{\delta}(X;\widehat{\bEta})$ and $\mu_{a,1}(X;\widehat{\bbeta}_{a})\;(a=0,1)$
be general semiparametric models for $\pi_{\delta}(X)$ and $\mu_{a,1}(X)\;(a=0,1)$,
respectively. Assume the following regularity conditions hold:
(C1) $\Vert\pi_{\delta}(X;\widehat{\bEta})-\pi_{\delta}(X)\Vert=o_{p}(1)$ 
and $\Vert\mu_{a,1}(X;\widehat{\bbeta}_{a})-\mu_{a,1}(X)\Vert=o_{p}(1)$,
for $a=0,1$; (C2) $\Vert\pi_{\delta}(X;\widehat{\bEta})-\pi_{\delta}(X)\Vert\sum_{a=0}^{1}\Vert\mu_{a,1}(X;\widehat{\bbeta}_{a})-\mu_{a,1}(X)\Vert=o_{p}(n^{-1/2})$.
Then $\htau^{\mathrm{ACW}}$ is consistent for $\tau_{0}$ and achieves
the semiparametric efficiency bound.
\end{theorem}

The semiparametric efficiency bound is attained as long as either
$\widehat{\bEta}$ or $(\widehat{\bbeta}_{0},\widehat{\bbeta}_{1})$
approximate the underlying sampling score model or the outcome models
well. (C1) states that we require that the posited
models be consistent. (C2) states that the
combined rate of convergence of the posited models is of $o_{p}(n^{-1/2})$.
In Section \ref{sec:Choice-of-g}, we construct such estimators using
the method of sieves, which satisfies (C1) and (C2) in Theorem~\ref{th:semi_nuisance} under regularity conditions.

For the locally efficient estimator $\htau^{\mathrm{ACW}}$, the variance
 estimator can be calculated as 
\begin{align}
  \widehat{V}\left(\htau^{\mathrm{ACW}}\right) = &  \sum_{i=1}^{N}
  \delta_i\widehat{q}_i\left[\frac{\widehat{V}\left\{ Y(1)|X_i,\delta_i\right\}}{\pi_{Ai}}
+\frac{\widehat{V}\left\{ Y(0)|X_i,\delta_i\right\}}{1-\pi_{Ai}}\right] \nonumber\\
 & +  \left\{ \sum_{i=1}^{N}\tildelta_{i}d_{i}\right\} ^{-2}\sum_{i=1}^{N}\tildelta_{i}d_{i}^2\left\{ \mu_{1,1}(X_{i};\widehat{\bbeta}_{1})-\mu_{0,1}(X_{i};\widehat{\bbeta}_{0}) - \htau^{\mathrm{ACW}}\right\} ^{2}, \label{eq:var_plugin}
\end{align}
where $\widehat{V}\left\{ Y(a)|X_i,\delta_i\right\}$ is a consistent estimator of ${V}\left\{ Y(a)|X_i,\delta_i\right\}$ for $a=0,1$. However, the plug-in variance estimator requires an additional consistent estimator of ${V}\left\{ Y(a)|X_i,\delta_i\right\}$, which can be difficult to obtain. 
The bootstrap variance estimator is more
 straightforward, 
 and it can accommodate
 situations where either one of the nuisance models is misspecified.
\vspace{-1ex}
\subsection{Semiparametric models by the method of sieves\label{sec:Choice-of-g}}

To overcome the model misspecification issue inherent to parametric
models, we consider the method of sieves, 
which allows flexible models for $\pi_{\delta}(X)$ and $\mu_{a,1}(X),\;(a=0,1)$. 
Although general sieves
such as Fourier series, splines, wavelets, and artificial neural networks
(\citealp{chen2007large}) are applicable,
the power series is most common. For a $p$-vector
of non-negative integers $\mathbb{\kappa}=(\kappa_{1},\ldots,\kappa_{p})$,
let $|\mathbf{\kappa}|=\sum_{l=1}^{p}\kappa_{l}$ and $X^{\mathbf{\kappa}}=\prod_{l=1}^{p}X_{l}^{\kappa_{l}}.$
Define a series $\{\mathbf{\kappa}(k):k=1,2,\ldots\}$ for all distinct
vectors of $\mathbf{\kappa}$ such that $|\mathbf{\kappa}(k)|\leq|\mathbf{\kappa}(k+1)|$.
Based on this series, we consider a $K$-vector $\bg(X)=\{g_{1}(X),\ldots,g_{K}(X)\}^{\top}=\{X^{\kappa(1)},\ldots,X^{\kappa(K)}\}^{\top}$.

In the presence of many sieve basis terms, variable selection is needed to include necessary terms and to exclude terms that could result in efficiency loss.
To guide selection, 
we attempt to compare the semiparametric efficiency bound $ V_{\mathrm{eff}}$ in Theorem \ref{th:semi_bd-1} with different types of covariates, which, however, does not lead to a definitive conclusion. Fortunately,  given that the OS sample is much larger than the trial sample,
 the first term of $V_{\mathrm{eff}}$ often dominates the second term. Thus, we focus on the comparison of the first term.

\begin{lemma}\label{lm:varcomparison}
Let $X^{C}$ be confounders, $X^{O}$ be outcome predictors, and $X^{I}$ be IVs, where $X^{C}$, $X^{O}$, $X^{I}$ are subsets of $\bg(X)$. Also, let  $X^{C + I} = X^{C} \cup X^{I}$. Define the first term of $V_{\mathrm{eff}}$ that depends on $X^{*}$ as
\begin{equation*}
    V_{1}^{*}=E\left[\frac{\delta}{\pi_{\delta}(X^{*})^{2}}\left\{ \frac{V\left\{ Y(1)|X^{*},\delta\right\} }{\pi_{A}}+\frac{V\left\{ Y(0)|X^{*},\delta\right\} }{1-\pi_{A}}\right\}\right],
\end{equation*}
where $^*$ can be $C, O, C + I$.
 Then, we have $V_{1}^{O}\le V_{1}^{C}\le V_{1}^{C+I}$.
\end{lemma}
\vspace{-2ex}
The proof of Lemma \ref{lm:varcomparison} is in Section \ref{subsec:varcomparison} of the supplementary material. Lemma \ref{lm:varcomparison} suggests that including outcome predictors and excluding IVs reduces $V_1$.
Thus, we propose a new basis selection procedure for sieves estimation and calibration adjusting for outcome predictors.  First, we  approximate $\mu_{a}(X)$ by the generalized sieve functions
 $\mu_{a}(X;\bbeta_{a}^{*})=m_{a}\{\bbeta_{a}^{*\top}\bg(X)\}$ with  $\bbeta_{a}^{*\top}=\arg\min_{\bbeta}E[\mu_{a}(X)-m_{a}\{\bbeta^{\top}\bg(X)\}]^{2}$ for $a=0,1$.
Since the number of basis functions controls the smoothness of sieves estimators, we can specify a sufficiently large $K$ as an
initial number and apply the penalization to regularize 
the variability of the estimators. 
Specifically,
let $\widehat{\bbeta}_{a}=\arg\min_{\bbeta\in\R^{K}}\sum_{i=1}^{N}({\delta}_{i}I(A_{i}=a)[Y_{i}-m_{a}\{\bbeta_{a}^{\top}\bg(X)\}]^{2}+\sum_{j=1}^{K}p_{\xi_{a}}(|\beta_{j}|)),
$
where $p_{\xi_{a}}(\cdot)$ is the smoothly clipped absolute deviation (SCAD) penalty function \citep{fan2001variable}, for $a=0,1$.
We choose the tuning parameters $\xi_{a}$ via  cross-validation.
Under certain regularity conditions 
$\widehat{\bbeta}_{a}$ satisfies the selection consistency and oracle
properties \citep[see][]{fan2001variable}. 

Second, we calibrate the sieve basis terms that are predictive of the outcome. Instead of calibrating the selected basis terms of the outcome predictors directly, we can construct the sieve basis for $\log\{\pi_{\delta}(X)\}$ by power series of the selected variables to capture the possible non-linear relationship between $\log\{\pi_{\delta}(X)\}$ and $X$. Then, we 
conduct penalized sieve estimation of $\pi_{\delta}(X)$ 
by solving the system of estimating equations (\ref{eq:lam}) with the SCAD penalty. By emphasizing the outcome predictors, our strategy provides guidance for variable selection for covariate balancing and efficient estimation.
Following \citet{shortreed2017outcome}, an alternative strategy of prioritizing outcome predictors is to use the outcome-adaptive Lasso for the sampling score model with the sieve basis of all covariates. This approach incorporates the outcome-covariate association to impose heavier penalties on the covariates that are not or weekly associated with outcome.

Coupling sieve approximation and variable selection, $\htau^{\mathrm{ACW}}$ with flexible approximations of the two nuisance functions achieves the root-$n$ consistency and the semiparametric efficiency bound under mild regularity conditions; see Section \ref{s:Conditions} of the supplementary material.

\subsection{Related works \label{sec:Related}}

There are several recent articles that focus on regularized balancing methods. 
\citet{athey2018approximate} proposed an approximate residual balancing method that first fits a regularized linear outcome model and then reweights the residuals to minimize covariate imbalance. Unlike our method, \citet{athey2018approximate} relied on the linear outcome model. \citet{ning2020robust} considered a doubly robust estimator that uses penalized maximum likelihood estimation of the nuisance functions and calibrates the estimated propensity score by balancing the selected outcome predictors. Similarly, \citet{tan2020model} and \citet{tan2020regularized} proposed a doubly robust estimator through regularized calibrated estimation using the expected calibrated loss function when fitting the propensity score model.
Unlike these approaches that estimate the propensity score that satisfies covariate balancing conditions, our method directly achieves the balance in the covariates through calibration weights, similar to \citet{chan2016globally}. Moreover, our approach uses the nonparametric sieve method which provides more robust estimation of the nuisance functions. The difference between our approach and \citet{chan2016globally} is that their approach enforces a three-way balance between the treated, the controls, and the combined data, whereas our method only requires a two-way balance between the RCT and the OS sample. 
Even though the three-way balancing approach is not necessary for generalizing trial results, it could be useful when generalizing observational results to a larger population. For example, in order to achieve double robustness in the observational setting, the CW estimator requires the three-way balance. It is analogous to the Covariate Balancing Propensity Score \citep[CBPS;][]{imai2014covariate} method which is doubly robust under the constant CATE whereas it requires the three-way balance to achieve double robustness under the heterogeneous CATE \citep{fan2021optimal}. 
Moreover, \citet{chan2016globally} did not solve the problem about which terms to calibrate, whereas we propose a principled approach for selecting calibration terms.

\citet{wang2020minimal} studied a class of weights that have minimum dispersion and showed that achieving approximate covariate balance corresponds to regularizing inverse probability weights, without explicitly involving the propensity score model. A special case is the Stable Balancing 
Weights method \citep[SBW; ][]{zubizarreta2015stable, chattopadhyay2020balancing} which finds weights with the minimum variance that achieves covariate balance approximately. The approximately balancing methods could be useful when the costs of balancing are too high, since they have the flexibility to trade bias for variance. Our strategy of handling large-dimensional calibration terms is different. We first reduce the number of calibration terms by selecting the outcome predictors and further use penalized estimating equations \citep{yang2020doubly} to obtain calibration weights. Both steps involve convex optimization with regularization, whose numerical and theoretical properties are well studied in the literature \citep{fan2001variable, johnson2008penalized}. Our solution for handling the large-dimension calibration terms is thus attractive in terms of feasibility and efficiency.

\subsection{ACW estimator when $Y$ and $A$ are available in OSs}

We consider another setting where we have access to additional 
information on $(A,Y)$ from the OS sample \citep[e.g.,][]{dahabreh2020benchmarking}. Most causal inference methods invoke the ``no unmeasured confounding''
assumption that $A$
is independent of the potential
outcomes given $X$ in the OS sample \citep[e.g.,][]{lu2019causal}.
To leverage the predictive power of the OS sample,
we assume generalizability of the outcome mean functions from the
RCT to the OS sample.

\begin{assumption} \label{assump: transp of rwd}
For $a=0,1$, $E(Y\mid X,A=a,\tilde{\delta}=1)=\mu_{a,1}(X)$.
\end{assumption}

Collectively, combining Assumptions \ref{a:consist}
- \ref{assump:design_wt} and \ref{assump: transp of rwd} leads to
generalizability of the CATE function: $E(Y\mid X,A=1,\tilde{\delta}=1)-E(Y\mid X,A=0,\tilde{\delta}=1)=\tau(X)$. The nuisance functions $\mu_{a,1}(X)$ ($a=0, 1$) in the ACW estimator $\htau^{\mathrm{ACW}}$ in (\ref{eq:taudc-1}) can be estimated
by the OS sample to further boost efficiency.
The indication is that the OS has no unmeasured confounding
on the mean difference measure conditional on $X$ \citep{vanderweele2012confounding}.
Assumption~\ref{assump: transp of rwd} is testable because it is
based only on the observed data. For example, one can use a likelihood
ratio test for testing a reduced model with the same outcome mean
model specification versus a full model with different model specifications
in the RCT and OS samples. 
Note that failure to reject this assumption does not ensure the whole
set of Assumptions \ref{a:consist} - \ref{assump:design_wt} and
\ref{assump: transp of rwd} holds; subject matter knowledge should be consulted, e.g., \citet{dahabreh2020benchmarking}.

\section{Simulation Study\label{sec:Simulation}}

We conduct simulation studies to evaluate the finite sample performances of the
proposed estimators. Table \ref{table:simulation} describes four simulation scenarios and twelve estimators to be compared, 
and Figure~\ref{fig:compareAll} displays the results with boxplots of the estimators. 
Details of the data generating process and numerical results are provided in Section \ref{s:simulation details} of the supplementary material.
It can be seen that the naive and IPSW estimators fail to adjust for the selection bias associated with the RCT sample. 
The SBW and CW estimators can correct the selection bias and are doubly robust, but they have larger variances than other doubly robust estimators.
In Scenario 3 when the outcome model is misspecified, the AIPSW estimator has a larger bias than other doubly robust estimators. This is because the AIPSW estimator is inflicted by the inverse probability of sampling weights, which, as shown in Scenario 1, results in the large finite-sample bias of the IPSW estimator. 
The ACW estimators 
do not involve weighting by the inverse probability of sampling and are more stable, thus we recommend the ACW estimators in practice.
The ACW-t(S$^{O}$) and ACW-b(S$^{O}$) are shown to be doubly robust and more efficient than other doubly robust estimators. The ACW-t, ACW-t(S), ACW-b, and ACW-b(S) are unbiased but show high variability, 
which could be due to the inclusion of IVs. 
In Scenario 4 where both outcome and sampling score models are misspecified, the ACW estimators focusing on outcome predictors,
i.e., ACW-t(S$^{O}$) and ACW-b(S$^{O}$) are still unbiased and efficient. Moreover, ACW-b(S$^{O}$) has smaller variance than ACW-t(S$^{O}$) by exploiting the predictive power from the OS
sample.

\begin{table}[h]
\caption{Simulation settings: description of four scenarios and estimators}
\footnotesize
\label{table:simulation} %
    \centering  
        \resizebox{\textwidth}{!}{%
    \begin{tabular}{ll}
    \toprule
    \textbf{Scenarios} & \textbf{Details} \tabularnewline \cmidrule(r){1-1} \cmidrule(l){2-2}
    1. O:C/S:C & Both outcome and sampling score models are correctly specified \tabularnewline
        \rowcolor{Gray} & The outcome model is correctly specified; the sampling score model is incorrectly specified by using $X^{*}$ \tabularnewline
        \rowcolor{Gray} \multirow{-2}{*}{2. O:C/S:W} & in the generative model \tabularnewline
    \multirow{2}{*}{3. O:W/S:C}& The outcome model is incorrectly specified by using $X^{*}$ in the generative model; the sampling score \tabularnewline
    & model is correctly specified  \tabularnewline
    \rowcolor{Gray} 4. O:W/S:W & Both outcome and sampling score models are incorrectly specified by using $X^{*}$ in the generative model \tabularnewline \midrule
    \textbf{Estimators} & \textbf{Details} \tabularnewline \cmidrule(r){1-1} \cmidrule(l){2-2}
    Naive & The difference in sample means of the two treatment groups in the RCT sample \tabularnewline
    \rowcolor{Gray} IPSW & The IPSW estimator with a logistic sampling score model \tabularnewline
    AIPSW & The AIPSW estimator with a logistic sampling score model \tabularnewline
    \rowcolor{Gray} & The AIPSW estimator using methods of sieve with $\bg(X)=\bg_{2}(X)$ for sampling score and outcome \tabularnewline
    \rowcolor{Gray} \multirow{-2}{*}{AIPSW(S)} & models based on the trial sample \tabularnewline
    \multirow{2}{*}{SBW} &  The IPSW estimator with SBW-1 weights \citep{chattopadhyay2020balancing} of the average treatment effect  \tabularnewline 
    & on the treated, with the OS being the treatment group and the RCT sample being the control group
    \tabularnewline
    \rowcolor{Gray} CW & The CW estimator defined by \eqref{eq:cbss0} with $\bg(X)=\bg_{1}(X)$ \tabularnewline
    \multirow{2}{*}{ACW-t} & The ACW estimator defined by \eqref{eq:taudc-1} with $\bg(X)=\bg_{1}(X)$ and the nuisance functions $\mu_{a}(X,1)$ and $\mu_{0}(X,1)$ \tabularnewline
    & are estimated based on the trial sample \tabularnewline
    \rowcolor{Gray} & The penalized ACW-t estimator using the method of sieves with $\bg(X)=\bg_{2}(X)$ for sampling score and  \tabularnewline
    \rowcolor{Gray} \multirow{-2}{*}{ACW-t(S)} & outcome models respectively \tabularnewline
     & The penalized ACW estimator using the method of sieves with $\bg(X)=\bg_{2}(X)$ for outcome models and \tabularnewline 
    \multirow{-2}{*}{ACW-t(S$^{O}$)}& construct the sieve basis for $\pi_{\delta}(X)$ by power series of the selected outcome predictors\tabularnewline 
    \rowcolor{Gray} & The ACW estimator defined by \eqref{eq:taudc-1} with $\bg(X)=\bg_{1}(X)$ and the nuisance functions $\mu_{1}(X)$ and $\mu_{0}(X)$ are \tabularnewline
    \rowcolor{Gray} \multirow{-2}{*}{ACW-b}& estimated based on both RCT and OS samples \tabularnewline
    \multirow{2}{*}{ACW-b(S)} & The penalized ACW-b estimator using the method of sieves with $\bg(X)=\bg_{2}(X)$ for sampling score and \tabularnewline
    & outcome models respectively \tabularnewline
    \rowcolor{Gray} & The penalized ACW-b estimator using the method of sieves with $\bg(X)=\bg_{2}(X)$ for outcome models and  \tabularnewline
    \rowcolor{Gray} \multirow{-2}{*}{ACW-b(S$^{O}$)} & construct the sieve basis for $\pi_{\delta}(X)$ by power series of the selected outcome predictors\tabularnewline 
    \bottomrule
    \end{tabular}
    }
\end{table}

\begin{figure}
\centerline{
\includegraphics[width=6in]{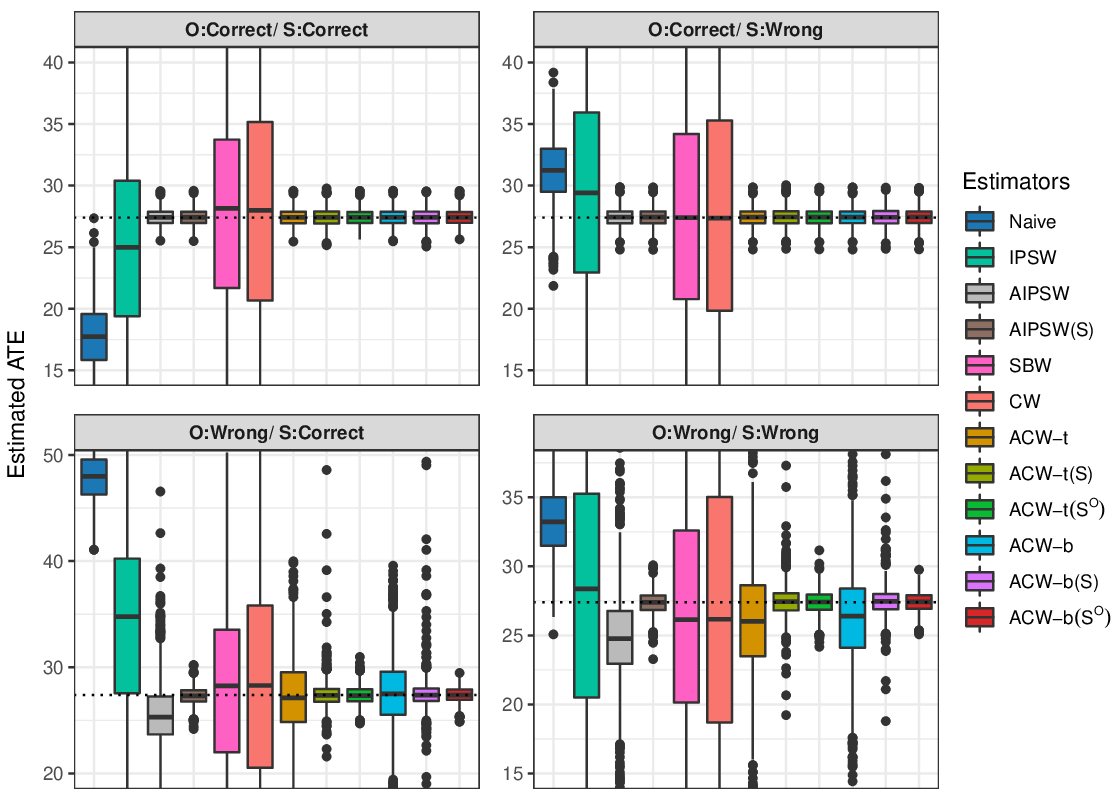}}
\caption{Boxplot of estimators under four model
specification scenarios, where a few outliers are removed for visualization. \label{fig:compareAll}}
\end{figure}

\newpage

\section{Real Data Application \label{sec:Real-data-application}}

We apply the proposed estimators to evaluate the effect of adjuvant
chemotherapy for early-stage resected non-small cell lung cancer (NSCLC).
Adjuvant chemotherapy for resected NSCLC is shown to be effective
in stages II and IIIA disease based on RCTs \citep{massarelli2003retrospective};
however, its utility in the early-stage disease remains unclear. Cancer
and Leukemia Group B (CALGB) 9633 is the only trial
designed specifically to evaluate the benefit of adjuvant chemotherapy over observation for stage IB NSCLC patients after surgery \citep{strauss2008adjuvant}.
Additional OS data for stage IB NSCLC patients were extracted from National Cancer Database (NCDB) with the same eligibility criteria as CALGB 9633. 
NCDB is a large joint project of the American Cancer Society and the American College of Surgeons,
and it captures 70\% of all cancers diagnosed in the US. It is designed
to be a registry, and there is no design weights associated with this
database \citep{jairam2019strengths}. 
As the extracted OS samples from NCDB were diagnosed between the years 2004--2016, and CALGB 9633 enrolled patients between the years 1996--2003, the patients of the two sources can be considered independent.

Table \ref{tab:Assumption-Justitifation} discuss the plausibility of the identification assumptions, and Table \ref{tab:Summary} (a) summarizes the baseline characteristics of the CALGB 9633 trial sample and the NCDB sample. The treatment indicator $A$ is coded as $1$ for
adjuvant chemotherapy and $0$ for on observation. The
outcome is the indicator of cancer recurrence within three years after
the surgery. 
The four covariates have been considered strong prognostic factors for disease recurrence after surgical resection for early NSCLC. As seen in Table  \ref{tab:Summary} (a), there are significant differences in the distribution of these covariates between the two data sources. Specifically, 
CALGB 9633 has a significantly higher percentage of male and younger ($<$ 70 years old) patients with smaller tumor size. While adjuvant chemotherapy is now recommended to stage IB NSCLC patients with a tumor size $>4$cm \citep{nccn},
it remains an important question whether adjuvant chemotherapy benefits
the general NSCLC patient population represented by NCDB, with a higher percentage of female and older age and larger tumor size. As these covariates are strong prognostic factors of disease recurrence and they may even be modifiers for the treatment effect of adjuvant chemotherapy, naive estimators based only on CALGB 9633 data will lead to biased quantification of the true treatment effect defined on the entire population of early-stage NSCLC patients.

We compare the proposed estimators with other ATE estimators, same as in the simulation studies. For
sieves estimators, the basis functions are the first and second-order
moments of the four covariates. We select a subsample using 1:10 matching based on the observed covariate and combine the RCT and matched OS samples for fitting outcome regression in the ACW-b(S) and ACW-b(S$^{O}$) methods. Bootstrap variance estimation is applied to estimate the standard errors. 

Table \ref{tab:Summary} (b) reports the results. The results indicate
that in the RCT sample, there is an $8.3
$ \% decrease in the risk of recurrence for adjuvant chemotherapy over
observation. The IPSW, AIPSW, AIPSW(S), SBW, ACW-t(S) and ACW-t(S$^{O}$) estimators, which utilized
OS covariate information, show a $9\%-14\%$ decrease
in the risk of recurrence. However, the causal effect is not significant
according to the 95\% confidence interval. By leveraging the predictive power of the OS sample, the ACW-b(S) and ACW-b(S$^{O}$) estimators give an estimate of $17\%$ risk decrease, which is significant at $0.05$ level. Moreover, the ACW-t(S$^{O}$) and ACW-b(S$^{O}$) estimators gain efficiency by focusing on outcome predictors, compared to ACW-t(S) and ACW-b(S).
All of the sampling score corrected
estimators have steeper declines in recurrence risk compared to the
naive estimator, which suggests that the causal risk difference
in the target population is larger than the one of the RCT sample,
i.e., the effect of adjuvant chemotherapy is more profound in the
real patient population.

\begin{table}
\caption{\label{tab:Assumption-Justitifation} Justification of the identification assumptions in the context of the CALGB 9633 trial and the NCDB sample.}
\resizebox{\textwidth}{!}{
\begin{tabular}{ll}
\toprule 
\textbf{Assumptions} & \textbf{Justifications} \tabularnewline
\cmidrule(r){1-1} \cmidrule(l){2-2}
\multirow{6}{*}{\ref{a:consist} Consistency} & The extracted OS samples are stage IB NSCLC patients who had surgery and then received\tabularnewline 
&  either adjuvant chemotherapy or on observation (i.e., no chemotherapy) and with age greater\tabularnewline
&  than 20. Like CALGB 9633 patients, they did not receive any of the neoadjuvant chemotherapy, \tabularnewline
& radiation therapy, induction therapy, immunotherapy, hormone therapy, transplant/endocrine \tabularnewline
&  procedures, or systemic treatment before their surgery. Thus, the same treatment or comparison \tabularnewline
& conditions were given in the same setting in both studies. \tabularnewline
\rowcolor{Gray} & The CALGB 9633 trial implemented treatment randomization and had good patient compliance \tabularnewline
\rowcolor{Gray} \multirow{-2}{*}{\ref{a:ign} Treatment ignorability and positivity} & \citep{strauss2008adjuvant}.\tabularnewline
\multirow{4}{*}{\ref{a:pos} Sampling ignorability and positivity} & The four covariates, gender, age, histology, and tumor size, have been considered strong prognostic  \tabularnewline 
& factors or disease recurrence after surgical resection for early NSCLC. The positivity condition \tabularnewline
& holds because the OS data for NSCLC stage IB patients were extracted from NCDB with the\tabularnewline
& same eligibility criteria as CALGB 9633.\tabularnewline
\rowcolor{Gray}  & The likelihood ratio test of a reduced model (i.e., a single logistic regression with the sieve basis \tabularnewline
\rowcolor{Gray} & for the combined sample) against a full model (i.e., two separate logistic regressions with the  \tabularnewline
\rowcolor{Gray} & sieve basis for the two samples) has a p-value of 0.09. If a conservative investigator uses 0.1 to \tabularnewline
\rowcolor{Gray}& determine the critical value, the investigator can choose estimators using only trial data, e.g.,\tabularnewline
\rowcolor{Gray} &  ACW-t(S) and ACW-t(S$^{O}$). On the other hand, if the investigator uses 0.05 to determine the  \tabularnewline
\rowcolor{Gray} \multirow{-6}{*}{ \parbox{6.5cm}{\ref{assump: transp of rwd} Generalizability of the outcome \\ \textcolor{Gray}{7} mean functions from the RCT \\ \textcolor{Gray}{7} sample to the OS sample} }& critical value, one can choose estimators using both data sources, i.e., ACW-b(S) and ACW-b(S$^{O}$).\tabularnewline
\bottomrule
\end{tabular}
}
\end{table}

\begin{table}
\caption{\label{tab:Summary} (a) Summary of baseline characteristics of the CALGB 9633 trial sample and the NCDB sample. (b) Point estimate, standard error, and 95\% percentile confidence interval of the causal risk difference between adjuvant
chemotherapy and observation based on the CALGB 9633 trial sample and the NCDB sample.}
\centering
\begin{tabular}{l}
(a) \\ 
\resizebox{0.7\textwidth}{!}{
\begin{tabular}{lcc}
\toprule 
 & \textbf{RCT: CALGB 9633} & \textbf{OS: NCDB}\tabularnewline
 & $\mathbf{n = 319}$ & $\mathbf{n = 15379}$ \tabularnewline
\midrule 
\textbf{Recurrence} ($Y$), $n (\%)$ & $ 79 \ (25)$ & $5060 \ (33)$\tabularnewline
\rowcolor{Gray}\textbf{Treatment} ($A$), $n (\%)$ & &\tabularnewline
\rowcolor{Gray}\ \ \ Adjuvant chemotherapy & $ 156 \ (49)$ & $4324 \ (28)$ \tabularnewline
\rowcolor{Gray}\ \ \ Observation &  $ 163 \ (51)$ & $11055 \ (72)$  \tabularnewline
\rowcolor{white} \textbf{Gender} ($X_1$), $n (\%) $ & & \tabularnewline
\ \ \ Male & $ 204 \ (64) $ & $ 8458 \ (55)$ \tabularnewline
\ \ \ Female & $115 \ (36) $ & $6921 \ (45)$  \tabularnewline
\rowcolor{Gray}\textbf{Age} ($X_2$), $mean \pm SD$ & $ 60.83 \pm 9.62 $ & $67.87 \pm 10.18 $ \tabularnewline
\rowcolor{white} \textbf{Histology} ($X_3$),  $n (\%) $ & & \tabularnewline
\ \ \ Squamous & $ 128 \ (40)$ & $ 5998 \ (39) $  \tabularnewline
\ \ \ Non-squamous & $ 191 \ (60)$ & $9381 \ (61)$  \tabularnewline
\rowcolor{Gray}\textbf{Tumor size} ($X_4$), $mean \pm SD$ & $ 4.6 \pm 2.08$ & $4.94 \pm 3.04 $  \tabularnewline
\bottomrule 
\end{tabular}
}\\ \\

(b)\\
\resizebox{0.7\textwidth}{!}{
\rowcolors{2}{Gray}{white}
\footnotesize
\begin{tabular}{cccc}
\toprule
 & $\widehat{\btau}$  & $\widehat{\bm{SE}}(\widehat{\btau})$ & \textbf{95\%  CI} \tabularnewline
\midrule
Naive & $-0.083$ & $0.044$ & $(-0.163, -0.018)$\tabularnewline
IPSW & $-0.088$ & $0.060$ & $(-0.211,0.019)$ \tabularnewline
AIPSW & $-0.088$ & $0.060$ & $(-0.187, 0.041)$ \tabularnewline
AIPSW(S) & $-0.106$ & $0.068$ & $(-0.233,0.020)$ \tabularnewline
SBW & $-0.090$ & $0.057$ & $(-0.187,0.017)$ \tabularnewline
CW & $-0.105$ & $0.058$ & $(-0.221, 0.000)$ \tabularnewline
ACW-t(S) & $-0.139$ & $0.106$ & $(-0.309, 0.041)$ \tabularnewline
ACW-t$\left(\mbox{S}^{O}\right)$ & $-0.122$ & $0.080$ & $(-0.237, 0.054)$ \tabularnewline
ACW-b(S) & $-0.174$ & $0.098$ & $(-0.360, -0.044)$ \tabularnewline
ACW-b$\left(\mbox{S}^{O}\right)$ & $-0.172$ & $0.088$ & $(-0.357, -0.050)$ \tabularnewline
\bottomrule 
\end{tabular}
}
\end{tabular}
\end{table}

\newpage

\section{Discussion \label{sec:Concluding remarks}}

In this paper, we have developed a new semiparametric framework to
evaluate the average treatment effects integrating the complementary
features of the RCTs and OSs under assumptions of
RCT randomization of treatment, generalizability of the CATE or the
outcome mean functions and positivity of trial participation. 
The proposed framework can be extended to the indirect comparison problem \citep[e.g.,][]{phillippo2018methods} under the transportability of CATEs, which we will pursue in the future.

In real data application, we assume that the RCT sample and the OS sample are independent
based on the study designs for the CALGB trial and the NCDB study.
In general, this assumption would be violated if there is a significant
overlapping of the two data sources, i.e., they involve the same subset
of patients. We note that the violation of this assumption would not
affect the unbiasedness of the estimators but variance estimation.
Recently, \citet{saegusa2019large} developed a new weighted empirical
process theory for merged data from potential overlapping sources.
This inference framework does not require identifying duplicated individuals
and therefore is attractive. In the future, we will extend this inference
framework to our general setting of combining RCTs and OSs.

We have focused on the setting when all relevant covariates in $X$
are captured in both RCTs and OSs. However, because OSs were 
not initially collected for research purposes,
some important covariates may not be available from the OS.
\citet{yang2019combining} developed integrative causal analyses
of the ATEs combining big main data with unmeasured confounders and
smaller validation data with a full set of confounders; however, they
assumed that the validation sample (i.e., the RCT sample in our context)
is representative of the target population. In the presence of unmeasured
covariates in the OSs, there may be lingering selection
biases after calibration on the measured covariates. The future work
will investigate the sensitivity to the unmeasured covariates 
\citep{nguyen2017sensitivity, yang2017sensitivity}.

\section*{Acknowledgments}
Yang is partially supported by the NSF DMS 1811245, NIH P01 CA142538, 1R01AG066883, and 1R01ES031651. Zeng is partially supported by GM124104 and MH117458. Wang is partially supported by NIH P01 CA142538 and 1R01AG066883. 

\bibliographystyle{Chicago}
\bibliography{ci1, ci2}

\newpage{} 
\begin{center}
\textbf{\Large{}Supplementary Material}{\Large{} }{\Large\par}
\par\end{center}

\pagenumbering{arabic} 
\renewcommand*{\thepage}{S\arabic{page}}

\setcounter{lemma}{0} 
\global\long\def\thelemma{\textup{S}\arabic{lemma}}%
\setcounter{equation}{0} 
\global\long\def\theequation{S\arabic{equation}}%
\setcounter{section}{0} 
\global\long\def\thesection{S\arabic{section}}%
\setcounter{assumption}{0} 
\global\long\def\theassumption{S\arabic{assumption}}%
\setcounter{theorem}{0} 
\global\long\def\thetheorem{S\arabic{theorem}}%
\setcounter{figure}{0} 
\global\long\def\thefigure{S\arabic{figure}}%
\setcounter{table}{0} 
\global\long\def\thetable{S\arabic{table}}%
\global\long\def\thecondition{S\arabic{condition}}%
 \setcounter{equation}{0}

\global\long\def\thesubsection{S\arabic{section}.\arabic{subsection}}%

The supplementary material is organized as the following. In Section \ref{s:gen_trans}, we discuss the subtle differences between generalizability and transportability. In Section~\ref{s:ipsw_est}, we specify the IPSW and AIPSW estimators and provide an identification formula based on inverse sampling score weighting. In Section~\ref{s:proof}, we provide the proofs of Theorems 1--4 and Lemma 1. Section~\ref{s:Conditions} provides regularity conditions for the sieves estimator. Section \ref{s:Simulation} provides simulation study details and additional simulation results.

\section{Generalizability and Transportability}\label{s:gen_trans}

The problems of generalizability \citep{cole2010generalizing,stuart2011use,hernan2011compound,tipton2013improving,o2014generalizing,stuart2015assessing,keiding2016perils,dahabreh2019extending,dahabreh2019generalizing} and transportability \citep{pearl2011transportability,rudolph2017robust, westreich2017transportability, josey2021transporting} aim at extending findings from a randomized controlled trial (RCT) to a target population \citep{dahabreh2020extending}. However, there are subtle differences between the problems of generalizability and transportability in terms of estimands and identification assumptions.The differences in generalizability and transportability are summarized in Table \ref{tab:gvst} and also illustrated using causal diagrams in Figure \ref{fig:diagram}. 

The goal of generalizability is to generalize results from a trial to its larger population, whereas the goal of transportability is to extend results from a trial to a different external population. In generalizability, one assumes that the triplet $\{X, Y(0), Y(1)\}$ in the target population follows $P\{X, Y(0), Y(1)\}$, and the observational sample ($\widetilde{\delta} = 1$) is representative of the target population. The trial sample ($\delta = 1$) is selected from the population according to $P(\delta = 1 \mid X)$. 
The generalizability problem tries to use the trial sample to draw conclusions for the population, leveraging the observational sample. Thus, the estimand of interest is the functional form of $P\{X, Y(0), Y(1)\}$, e.g., the average treatment effect (ATE), $E\{Y(1)-Y(0)\}$. On the other hand, in transportability, we have two study samples following $P\{X, Y(0), Y(1) \mid \delta = 1\}$ and $P\{X, Y(0), Y(1) \mid \widetilde{\delta} = 1\}$ respectively, where we try to transport some features of $P\{X, Y(0), Y(1) \mid \delta = 1 \}$ to $P\{X, Y(0), Y(1) \mid \widetilde{\delta} = 1\}$. Thus, the estimand of interest is the functional form of $P\{X, Y(0), Y(1) \mid \widetilde{\delta} = 1\}$, e.g., the target population average treatment effect \citep[TATE; ][]{josey2021transporting}, $E\{Y(1)-Y(0)\mid \widetilde{\delta} = 1\}$. In transportability, we do not necessarily have to define $P\{X, Y(0), Y(1)\}$, and we can use the conditional odds, ${P\{\widetilde\delta = 1 \mid X , (\delta=1 \textrm{ or } \widetilde\delta=1)\} }/{P\{ {\delta} = 1 \mid X, (\delta=1\textrm{ or }\widetilde\delta=1)\} }$, without defining $P(\delta = 1 \mid X)$ to estimate the TATE.
Under generalizability, $P\{X, Y(0), Y(1) \mid \widetilde{\delta} = 1\} = P\{X, Y(0), Y(1)\}$, i.e., both trial and observational samples are drawn from the broader trial population, whereas under transportability, $P\{X, Y(0), Y(1) \mid \widetilde{\delta} = 1\} \neq P\{X, Y(0), Y(1)\}$, i.e., the observational sample is drawn from the external population. We call the population which we want to make inferences about as the target population.

Identification assumptions (i) - (ii) are common for both generalizability and transportability and hold for well-defined RCTs in general. The key differences between generalizability and transportability are identification assumptions (iii) - (iv), which are often needed to extend RCT findings to the target population. Unlike transportability where these assumptions are often needed for the target population with $\widetilde{\delta} = 1$, for generalizability, these assumptions are needed for all $x$ such that $P(X = x) > 0$, assuming that the trial sample is drawn from the target population. Moreover, the positivity assumption of the trial participation (iv) requires the probability of sampling to be bounded away from 0 for generalizability, whereas bounded away from 0 and 1 for transportability \citep{degtiar2021review}. These suggest that in transportability analysis, variables separating the trial sample from the target population should be excluded from the measure exchangeability/mean exchangeability/ignorability assumption over $\delta$ \citep{tipton2013improving,dahabreh2020extending}.

The mean exchangeability over treatment assignment (i) is weaker than the ignorability assumption on treatment assignment in (i$^{*}$). Similarly, the measure exchangeability assumption (iii) is a weaker version of the mean exchangeability assumption (iii$^{*}$), which is weaker than the ignorability assumption (iii$^{**}$) \citep{dahabreh2019generalizing,dahabreh2020extending}. Under assumption (iii), the ATE or TATE are identifiable, but not the potential outcome means $E\{Y(a)\}$ or $E\{Y(a) \mid \widetilde{\delta} = 1\}$, $a = 0, 1$. Under assumptions (iii$^{*}$) and  (iii$^{**}$), $E\{Y(a)\}$ and thus the ATE/TATE are identifiable. If the scientific interest lies in the distributions or the means of the potential outcomes, stronger assumptions are required.

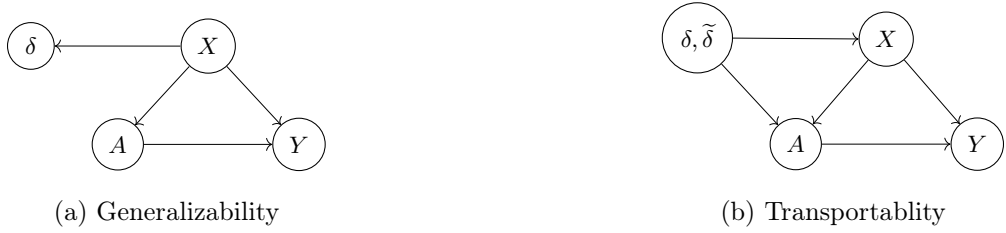
\begin{figure}[ht]
    \footnotesize
    \begin{subfigure}[b]{0.47\textwidth}
        \centering
        \begin{tikzcd}[column sep=1.5em, cells={nodes={draw, circle}}]
	    \delta  && X \arrow[left = 10]{ll} \arrow{dl} \arrow{dr} \\ 
	    & A \arrow{rr} & & Y 
        \end{tikzcd}
        \caption{Generalizability}
        \label{subfig:gen}
    \end{subfigure}
    \hfill
    \begin{subfigure}[b]{0.47\textwidth}
        \centering
        \begin{tikzcd}[column sep=1.5em, cells={nodes={draw, circle}}]
	        \delta, \widetilde{\delta}  \arrow{dr} \arrow{rr}  
	        & & X  \arrow{dl} \arrow{dr} \\ 
	       & A \arrow{rr} & & Y 
        \end{tikzcd}
        \caption{Transportablity}
        \label{subfig:tran}
    \end{subfigure}    
    \caption{Illustrations of causal diagrams in the problems of generalizability and transportability. 
    In (a), the RCT sample is subject to selection bias (indicated by $X$ pointing to $\delta$). 
    In (b), the two populations differ by covariate distributions (indicated by $\delta$ and $\widetilde{\delta}$  pointing to $X$) and the two populations differ in their treatment assignment mechanism ($\delta$ and $\widetilde{\delta}$  pointing to $A$). (b) is modified from \citet{pearl2011transportability}.
    }
    \label{fig:diagram}    
\end{figure}

\begin{table}[ht]
   \caption{Differences of the problems of Generalizability (G) and Transportability (T)}
    \label{tab:gvst}
    \scriptsize
    \centering
    \resizebox{\textwidth}{!}{%
    \begin{tabular}{|c|c|lc|}
        \hline
        \multirow{11}{*}{\textbf{G}} & \textbf{Goal} & \multicolumn{2}{c|}{Generalize results from a trial to \textit{its larger population}} \\\cline{2-4}
        & \textbf{Samples, populations} & \multicolumn{2}{c|}{A trial sample and an observational sample from a target population} \\\cline{2-4}
         & \textbf{Study design} & \multicolumn{2}{c|}{Nested design, non-nested design} \\\cline{2-4}
         & \textbf{Estimand} & \multicolumn{1}{c}{ATE} & $E \left\{Y(1)-Y(0) \right\}$ \\\cline{2-4}
         & \multirow{7}{2cm}{\textbf{Identification assumptions}} & (i) Mean exchangeability over $A$ & $E\{Y(a)\mid X,\delta=1,A=a\}=E\{Y(a)\mid X,\delta=1\}$ for $a = 0, 1$\\
         & & \cellcolor{Gray}(i$^{*}$) Ignorability on $A$ & \cellcolor{Gray} $Y(a)\indep A\mid(X,\delta=1)$ for $a=0,1$ \\
         & & (ii) Positivity of $A$ &  $0<P(A = a \mid X = x,\delta = 1)<1 ~ \forall x \mbox{ s.t. } P(X = x | \delta = 1) > 0 $\\
         & & \cellcolor{Gray}(iii) Measure exchangeability over $\delta$ & \cellcolor{Gray}  $E\{Y(1) - Y(0) \mid X,\delta=1\}=E\{Y(1) - Y(0) \mid X\}$ \\
         & & (iii$^{*}$) Mean exchangeability over $\delta$ & $E\{Y(a)\mid X,\delta=1\}=E\{Y(a)\mid X\}$ for $a = 0, 1$\\
         & & \cellcolor{Gray}(iii$^{**}$) Ignorability on $\delta$ & \cellcolor{Gray}  $Y(a)\indep\delta\mid X$ for $a=0,1$ \\
         & & (iv) Positivity of $\delta$ & $P(\delta = 1 \mid X = x)>0$ with probability 1 \\\hhline{|=|=|==|}
        \multirow{14}{*}{\textbf{T}} & \textbf{Goal} & \multicolumn{2}{c|}{Extend results from a trial to a \textit{different external population}}\\\cline{2-4}
        &  \multirow{2}{*}{\textbf{Samples, populations}} & \multicolumn{2}{c|}{A trial sample from a trial population and an observational sample from a target population}\\
        & & \multicolumn{2}{c|}{where the trial and the target population are not identical}\\\cline{2-4}
         & \textbf{Study design} & \multicolumn{2}{c|}{Non-nested design} \\\cline{2-4}
         & \textbf{Estimand} & \multicolumn{1}{c}{TATE} &  $E \left\{Y(1)-Y(0)\mid \widetilde{\delta} = 1 \right\}$ \\ \cline{2-4}
         & \multirow{9}{2cm}{\textbf{Identification assumptions}} & (i) Mean exchangeability over $A$ & same as in G(i)\\
         & & \cellcolor{Gray}(i$^{*}$) Ignorability on $A$ & \cellcolor{Gray} same as in G(i$^{*}$) \\
         & & (ii) Positivity of $A$ & same as in G(ii) \\
         & & \cellcolor{Gray}  & \cellcolor{Gray} $E\left\{Y(1) - Y(0)\mid X = x,\delta=1\right\}=E\left\{Y(1) - Y(0)\mid X = x, \widetilde{\delta} = 1\right\}$\\
         & &  \cellcolor{Gray}\multirow{-2}{*}{(iii) Measure exchangeability over $\delta$}& \cellcolor{Gray} $~\forall x \mbox{ s.t. } P\left(X = x \mid \widetilde{\delta} = 1\right) > 0 $ \\         
         & & & $E\left\{Y(a)\mid X = x,\delta=1\right\}=E\left\{Y(a)\mid X = x, \widetilde{\delta} = 1\right\}$\\
         & & \multirow{-2}{*}{(iii$^{*}$) Mean exchangeability over $\delta$} & $~\forall x \mbox{ s.t. } P\left(X = x \mid \widetilde{\delta} = 1\right) > 0 $  and $a=0,1$ \\
         & &  \cellcolor{Gray}(iii$^{**}$) Ignorability on $\delta$ &  \cellcolor{Gray}$Y(a)\indep\delta\mid X = x ~\forall x \mbox{ s.t. }  P\left(X = x \mid \widetilde{\delta} = 1\right) > 0$  and $a=0,1$\\
         & & (iv) Positivity of $\delta$ &  $0 < P(\delta = 1 \mid X = x) < 1 ~\forall x \mbox{ s.t. } P\left(X = x \mid \widetilde{\delta} = 1\right) > 0 $ \\\hline
    \end{tabular}
 }
\end{table}

\newpage

\section{IPSW and AIPSW}\label{s:ipsw_est}
The IPSW estimator of the ATE is
\begin{equation}
\hat{\tau}^{\ipsw}=\frac{\sum_{i=1}^{n}\pi_{\delta}(X_{i};\widehat{\bEta})^{-1}A_{i}Y_{i}}{\sum_{i=1}^{n}\pi_{\delta}(X_{i};\widehat{\bEta})^{-1}A_{i}}-\frac{\sum_{i=1}^{n}\pi_{\delta}(X_{i};\widehat{\bEta})^{-1}(1-A_{i})Y_{i}}{\sum_{i=1}^{n}\pi_{\delta}(X_{i};\widehat{\bEta})^{-1}(1-A_{i})}.\label{eq:ipsw}
\end{equation} 

The augmented inverse probability weighting estimator (AIPSW) has been proposed to improve it by employing both the sampling
score and outcome regression
\begin{multline}\label{eq:aipw}
\hat{\tau}^{\aipsw}=\frac{\sum_{i=1}^{n}\pi_{\delta}(X_{i};\widehat{\bEta})^{-1}A_{i}\{Y_{i}-\widehat{\mu}_{1,1}(X)\}}{\sum_{i=1}^{n}\pi_{\delta}(X_{i};\widehat{\bEta})^{-1}A_{i}} - \\ 
 \frac{\sum_{i=1}^{n}\pi_{\delta}(X_{i};\widehat{\bEta})^{-1}(1-A_{i})\{Y_{i}-\widehat{\mu}_{0,1}(X_{i})\}}{\sum_{i=1}^{n}\pi_{\delta}(X_{i};\widehat{\bEta})^{-1}(1-A_{i})} + \\
\frac{1}{m}\sum_{i=n+1}^{n+m}\{\widehat{\mu}_{1,1}(X_{i})-\widehat{\mu}_{1,1}(X_{i})\}.
\end{multline}
\subsection{Identification of IPSW estimator}
We provide another
identification formula based on IPSW.
We first show
that $\pi_{\delta}(X)$ is identifiable up to a constant $P(\delta=1)$
based on 
\[
\pi_{\delta}(X)=P(\tildelta=1\mid X)\frac{f(X\mid\delta=1)}{f(X\mid\tildelta=1)}\frac{P(\delta=1)}{P(\tildelta=1)}\coloneqq\widetilde{\pi}_{\delta}^{c}(X)\frac{P(\delta=1)}{P(\tildelta=1)},
\]
where we assume that $P(\tildelta=1\vert X)$ is known by design,
$f(X\mid\delta=1)$ and $f(X\mid\tildelta=1)$ are observed data distributions,
but $P(\delta=1)/P(\tildelta=1)$ is identifiable without the knowledge
of the sizes of the underlying populations. Nonetheless, the ATE can
be identified based on
\[
\tau_{0}=E\left[\frac{\delta\widetilde{\pi}_{\delta}^{c}(X)^{-1}AY/\pi_{A}(X)}{\delta\widetilde{\pi}_{\delta}^{c}(X)^{-1}A/\pi_{A}(X)}-\frac{\delta\widetilde{\pi}_{\delta}^{c}(X)^{-1}(1-A)Y/\{1-\pi_{A}(X)\}}{\delta\widetilde{\pi}_{\delta}^{c}(X)^{-1}(1-A)/\{1-\pi_{A}(X)\}}\right].
\]

\section{Proofs}\label{s:proof}

\subsection{Proof of Theorem~\ref{th:consistency1}\label{subsec:Proof-cw}}

\subsubsection*{Proof of the double robustness of the calibration weighting estimator}

Let $\mathbf{\boldsymbol{\mu}}_{\bg0}=E\left\{ \bg(X)\right\} $,
$\bar{\bg}_{0}=\bg(X)-\boldsymbol{\mu}_{\bg0}$. To use the M-estimator
theory \citep{boos2013essential}, we write (\ref{eq:lam}) as the
following estimating equations 
\begin{align}
\frac{1}{N}\sum_{i=1}^{N}C(X_{i},\widetilde{\delta}_{i};\bm{\mu_{g}}) & =\frac{1}{N}\sum_{i=1}^{N}\widetilde{\delta}_{i}d_{i}\{\bg(X_{i})-\bm{\mu_{g}}\}=0,\label{ee:1}\\
\frac{1}{N}\sum_{i=1}^{N}\zeta(X_{i},\delta_{i};\blam,\bm{\mu_{g}}) & =\frac{1}{N}\sum_{i=1}^{N}\delta_{i}\exp\left\{ \blam^{\top}\bg(X_{i})\right\} \left\{ \bg(X_{i})-\bm{\mu_{g}}\right\} =0.\label{ee:2}
\end{align}
First consider the case where Assumption \ref{asump:logit linear}
holds, we have $\pi_{\delta}(X)=\exp\{\bEta_{0}^{\top}\bg(X)\}$.
Notice that $\mathbf{\boldsymbol{\mathbf{\boldsymbol{\mu}}_{\bg0}}}$
is the solution to $E\{C(X;\mathbf{\boldsymbol{\mu}}_{\bg})\}=0$.
Taking expectation on the left hand side of \eqref{ee:2} with $\mathbf{\boldsymbol{\mu}}_{\bg}=\mathbf{\boldsymbol{\mu}}_{\bg0}$
leads to 
\[
E\{\zeta(X,\delta;\blam,\mathbf{\boldsymbol{\mu}}_{\bg0})\}=E\{E(\zeta(X,\delta;\blam,\mathbf{\boldsymbol{\mu}}_{\bg0})\vert X)\}=E\left(\pi_{\delta}(X)\exp\{\blam^{\top}\bg(X)\}\left[\bg(X)-E\{\bg(X)\}\right]\right).
\]
For the above conditional expectation to be zero, one needs $\pi_{\delta}(X)\exp\{\blam^{\top}\bg(X)\}$
to be a constant. As $\pi_{\delta}(X)=\exp\{\bEta_{0}^{\T}\bg(X)\}$,
we have $\pi_{\delta}(X)\exp\{\blam^{\top}\bg(X)\}=\exp\{(\bEta_{0}+\blam)^{\top}\bg(X)\}$.
Thus $\blam=-\bEta_{0}$ makes \eqref{ee:2} a system of unbiased
estimating equations. We point out that denominator in $\widehat{q}_{i}$
is an estimator of the population size $N$, i.e., 
\begin{align*}
\frac{1}{N}\sum_{i=1}^{n}\exp\{\blam^{\top}\bg(X_{i})\} & =\frac{1}{N}\sum_{i=1}^{N}\delta_{i}\exp\{\blam^{\top}\bg(X_{i})\}\\
 & =\frac{1}{N}\sum_{i=1}^{N}\delta_{i}\exp\left\{ -\bEta_{0}^{\top}\bg(X_{i})\right\} +O_{p}(n^{-1/2}N^{-1})\\
 & =1+O_{p}(N^{-1/2})+O_{p}(n^{-1/2}N^{-1})\\
 & =1+o_{p}(1).
\end{align*}
Therefore, 
\begin{equation}
\widehat{q}_{i}=q(X_{i};\widehat{\blam})=\frac{\exp\left\{ \widehat{\blam}^{\top}\bg(X_{i})\right\} }{\sum_{i=1}^{n}\exp\left\{ \widehat{\blam}^{\top}\bg(X_{i})\right\} }=\text{\ensuremath{\frac{1}{N}}}\frac{1}{\pi_{\delta}(X_{i};\bEta_{0})}+O_{p}(n^{-1/2}N^{-1});\label{eq:qi}
\end{equation}
i.e., $\widehat{q}_{i}\left\{ N\pi_{\delta}(X_{i};\bEta_{0})\right\} \rightarrow1$
as $n\rightarrow\infty$. Based on \eqref{eq:qi}, we have 
\begin{align}
\htau^{\mathrm{CW}} & =\sum_{i=1}^{N}\widehat{q}_{i}\delta_{i}\left\{ \frac{A_{i}Y_{i}}{\pi_{Ai}}-\frac{(1-A_{i})Y_{i}}{1-\pi_{Ai}}\right\} \label{eq:taucw0_consist}\\
 & =\frac{1}{N}\sum_{i=1}^{N}\frac{\delta_{i}}{\pi_{\delta}(X_{i};\bEta_{0})}\left\{ \frac{A_{i}Y_{i}}{\pi_{Ai}}-\frac{(1-A_{i})Y_{i}}{1-\pi_{Ai}}\right\} =\tau_{0}+O_{p}(N^{-1/2})+O_{p}(n^{-1/2})\nonumber \\
 & =\tau_{0}+o_{p}(1).\nonumber 
\end{align}
Therefore, $\htau^{\mathrm{CW}}$ is consistent for $\tau_{0}$.

Now consider the case where Assumption \ref{assump: outcome linear}
holds. Then we have

\begin{align*}
E\left(\sum_{i=1}^{n}\widehat{q}_{i}\left\{ \frac{A_{i}Y_{i}}{\pi_{Ai}}-\frac{(1-A_{i})Y_{i}}{1-\pi_{Ai}}\right\} \right) & =E\left[\sum_{i=1}^{n}\widehat{q}_{i}E\left\{ \frac{A_{i}Y_{i}}{\pi_{Ai}}-\frac{(1-A_{i})Y_{i}}{1-\pi_{Ai}}\mid X_{i},\delta_{i}=1\right\} \right]\\
 & =E\left[\sum_{i=1}^{n}\widehat{q}_{i}E\left\{ Y(1)-Y(0)\mid X_{i},\delta_{i}=1\right\} \right]\\
 & =\boldsymbol{\gamma}_{0}^{\top}E\left\{ \sum_{i=1}^{n}\widehat{q}_{i}\bg(X_{i})\right\} =\boldsymbol{\gamma}_{0}^{\top}E\left\{ \frac{1}{N}\sum_{i=1}^{N}\widetilde{\delta}_{i}d_{i}\bg(X_{i})\right\} \\
 & =E\left\{ \boldsymbol{\gamma}_{0}^{\top}\bg(X)\right\} =\tau_{0},
\end{align*}
where the equation on the third line is obtained by the balancing
constraint \eqref{eq:calibration constraints}. Under regularity conditions
for unbiased M-estimators, $\htau^{\mathrm{CW}}$ is consistent for
$\tau_{0}$.

We thus conclude the double robustness of $\htau^{\mathrm{CW}}$.

\subsubsection*{Proof of the asymptotic variance for the calibration weighting estimator}

We derive the asymptotic variance of $\htau^{\mathrm{CW}}$ under
Assumption \ref{assump: outcome linear} and \ref{asump:logit linear}
to facilitate the efficiency comparison of $\htau^{\mathrm{CW}}$
and $\htau^{\mathrm{ACW}}$.

Let $\btheta=(\bm{\mu_{g}}^{\top},\blam^{\top},\pi_{A},\tau)^{\top}$
to denote the vector of all parameters. The estimating function for
$\btheta$ is

\[
\bpsi(X,A,Y,\delta,\tildelta;\btheta)=\begin{pmatrix}C(X,\widetilde{\delta};\bm{\mu_{g}})\\
\zeta(X,\delta;\blam,\bm{\mu_{g}})\\
h(X,A,\delta;\blam,\pi_{A})\\
t(X,A,Y,\delta;\blam,\pi_{A},\tau)
\end{pmatrix},
\]
where $C(X,\widetilde{\delta};\bm{\mu_{g}})$ and $\zeta(X,\delta;\blam,\bm{\mu_{g}})$
are given in (\ref{ee:1}), (\ref{ee:2}), respectively, and 
\begin{align*}
h(X,A,\delta;\blam,\pi_{A}) & =\delta\exp\{\blam^{\top}\bg(X)\}(A-\pi_{A}),\\
t(A,X,Y,\delta;\blam,\pi_{A},\tau) & =\delta\exp\{\blam^{\top}\bg(X)\}\left\{ \frac{AY}{\pi_{A}}-\frac{(1-A)Y}{1-\pi_{A}}-\tau\right\} .
\end{align*}
Then $\widehat{\btheta}=(\widehat{\boldsymbol{\mu}}_{\bg}^{\top},\widehat{\blam}^{\top},\hat{\pi}_{A},\htau^{\mathrm{CW1}})^{\top}$
solves the joint estimating equation 
\[
\frac{1}{N}\sum_{i=1}^{N}\bpsi(X_{i},A_{i},Y_{i},\delta_{i};\btheta)=\bm{0}.
\]
Under standard regularity conditions in the M-estimator theory, we
have 
\[
\widehat{\btheta}-\btheta_{0}=\frac{1}{N}\sum_{i=1}^{N}\bpsi(X_{i},A_{i},Y_{i},\delta_{i};\btheta_{0})+o_{p}(N^{-1/2}),
\]
where $A(\btheta_{0})=E\left\{ -\nabla_{\btheta_{0}}\bpsi(\btheta_{0})\right\} $,
and $\btheta_{0}=(\boldsymbol{\mu}_{\bg0}^{\top},-\bEta_{0}^{\top},\pi_{A},\tau_{0})^{\top}$.
The asymptotic variance of $N^{-1/2}(\widehat{\btheta}-\btheta_{0})$
is $A^{-1}(\btheta_{0})B(\btheta_{0})A^{-1}(\btheta_{0})^{\top}$,
where $B(\btheta_{0})=E\left\{ \bpsi(\btheta_{0})\bpsi(\btheta_{0})^{\top}\right\} $.

To further express the asymptotic variance, we denote $q_{0}=q_{0}(X)=\exp\{-{\bEta_{0}}^{\top}\bg(X)\}$
and $\bar{\tau}(Y,A)=\left\{ AY/\pi_{A}-(1-A)Y/(1-\pi_{A})-\tau_{0}\right\} $.
Note that $E\{\bar{\tau}(Y,A)\vert X,\delta=1\}=\tau(X)-\tau_{0}$
and $E\left(\widetilde{\delta}d\right)=1$. Under Assumption \ref{asump:logit linear}
and \ref{assump: outcome linear}, $\text{\ensuremath{\pi_{\delta}(X)q_{0}(X)}\ensuremath{=1}}$
and $\tau(X)-\tau_{0}=\boldsymbol{\gamma}_{0}^{\top}\bar{\bg}_{0}$.
In the following derivation we use $\Rightarrow$ to indicate equality
when both Assumption \ref{assump: outcome linear} and \ref{asump:logit linear}
hold.

Using iterated expectation, we have 
\begin{align*}
A(\btheta_{0}) & =E\left\{ -\nabla_{\btheta_{0}}\bpsi(\btheta_{0})\right\} \\
 & =E\begin{pmatrix}\widetilde{\delta}dI_{K} & \bm{0}_{K\times K} & \bm{0}_{K\times1} & \bm{0}_{K\times1}\\
\delta q_{0}I_{K} & -\delta q_{0}\bar{\bg}\bg^{\top} & \bm{0}_{K\times1} & \bm{0}_{K\times1}\\
\bm{0}_{1\times K} & -\delta q_{0}(A-\pi_{A})\bg^{\top} & \delta q_{0} & 0\\
\bm{0}_{1\times K} & -\delta q_{0}\bar{\tau}(Y,A)\bg^{\top} & \delta q_{0}\left\{ \frac{AY}{\pi_{A}^{2}}+\frac{(1-A)Y}{(1-\pi_{A})^{2}}\right\}  & \delta q_{0}
\end{pmatrix}\\
 & =\begin{pmatrix}I_{K} & \bm{0}_{K\times K} & \bm{0}_{K\times1} & \bm{0}_{K\times1}\\
E\left(\delta q_{0}\right)I_{K} & -E\{\delta q_{0}(\bg-\bm{\mu}_{\bg0})\bg^{\top}\} & \bm{0}_{K\times1} & \bm{0}_{K\times1}\\
\bm{0}_{1\times K} & \bm{0}_{1\times K} & E\left(\delta q_{0}\right) & 0\\
\bm{0}_{1\times K} & -E\left[\pi_{\delta}(X)q_{0}\left\{ \tau(X)-\tau_{0}\right\} \bg^{\top}\right] & E\left[\delta q_{0}\left\{ \frac{AY}{\pi_{A}^{2}}+\frac{(1-A)Y}{(1-\pi_{A})^{2}}\right\} \right] & E\left(\delta q_{0}\right)
\end{pmatrix}.
\end{align*}
By block matrix inversion, 
\[
A(\btheta_{0})^{-1}=\begin{pmatrix}I_{K} & \bm{0}_{K\times K} & \bm{0}_{K\times1} & \bm{0}_{K\times1}\\
E\left(\delta q_{0}\bar{\bg}_{0}\bg^{\top}\right){}^{-1}E(\delta q_{0}) & -E\left(\delta q_{0}\bar{\bg_{0}}\bg^{\top}\right){}^{-1} & \bm{0}_{K\times1} & \bm{0}_{K\times1}\\
\bm{0}_{1\times K} & \bm{0}_{1\times K} & E(\delta q_{0})^{-1} & 0\\
A_{41} & A_{42} & A_{43} & A_{44}
\end{pmatrix},
\]
where 
\begin{align*}
A_{41} & =E\left[\delta q_{0}(X)\left\{ \tau(X)-\tau_{0}\right\} \bg^{\top}\right]E\left\{ \delta q_{0}(X)\bar{\bg}_{0}\bg^{\top}\right\} {}^{-1}\Rightarrow\boldsymbol{\gamma}_{0}^{\top}I_{K},\\
A_{42} & =-E\left\{ \delta q_{0}(X)\right\} {}^{-1}E\left[\pi_{\delta}(X)q_{0}(X)\left\{ \tau(X)-\tau_{0}\right\} \bg^{\top}\right]E\left\{ \delta q_{0}(X)\bar{\bg}_{0}\bg^{\top}\right\} ^{-1}\Rightarrow-\boldsymbol{\gamma}_{0}^{\top}I_{K},\\
A_{43} & =-E\left\{ \delta q_{0}(X)\right\} ^{-2}E\left[\delta q_{0}(X)\left\{ \frac{AY}{\pi_{A}^{2}}+\frac{(1-A)Y}{(1-\pi_{A})^{2}}\right\} \right]\Rightarrow-E\left[\delta q_{0}(X)\left\{ \frac{Y(1)}{\pi_{A}}+\frac{Y(0)}{1-\pi_{A}}\right\} \right],\\
A_{44} & =E\left\{ \delta q_{0}(X)\right\} {}^{-1}\Rightarrow1.
\end{align*}
Taking iterated expectation again, we have 
\begin{align*}
B(\btheta_{0}) & =E\left\{ \bpsi(\btheta_{0})\bpsi(\btheta_{0})^{\top}\right\} \\
 & =E\begin{pmatrix}\widetilde{\delta}d^{2}\bar{\bg}_{0}\bar{\bg}_{0}^{\top} & \bm{0}_{K\times K} & \bm{0}_{K\times1} & \bm{0}_{K\times1}\\
\bm{0}_{K\times K} & \delta q_{0}(X)^{2}\bar{\bg}_{0}\bar{\bg}_{0}^{\top} & \delta q_{0}(X)^{2}(A-\pi_{A})\bar{\bg}_{0} & \delta q_{0}(X)^{2}\bar{\tau}(Y,A)\bar{\bg}_{0}\\
\bm{0}_{1\times K} & \delta q_{0}(X)^{2}(A-\pi_{A})\bar{\bg}_{0}^{\top} & \delta q_{0}(X)^{2}(A-\pi_{A})^{2} & \delta q_{0}(X)^{2}\bar{\tau}(Y,A)(A-\pi_{A})\\
\bm{0}_{1\times K} & \delta q_{0}(X)^{2}\bar{\tau}(Y,A)\bar{\bg}_{0}^{\top} & \delta q_{0}(X)^{2}\bar{\tau}(Y,A)(A-\pi_{A}) & \delta q_{0}(X)^{2}\bar{\tau}(Y,A)^{2}
\end{pmatrix}\\
 & =\begin{pmatrix}B_{11} & \bm{0}_{K\times K} & \bm{0}_{K\times1} & \bm{0}_{K\times1}\\
\bm{0}_{K\times K} & B_{22} & \bm{0}_{K\times1} & B_{24}\\
\bm{0}_{1\times K} & \bm{0}_{1\times K} & B_{33} & B_{34}\\
\bm{0}_{1\times K} & B_{24}^{\top} & B_{34} & B_{44}
\end{pmatrix},
\end{align*}
where 
\begin{align*}
B_{11} & =E\left(\widetilde{\delta}d^{2}\bar{\bg}_{0}\bar{\bg}_{0}^{\top}\right),\\
B_{22} & =E\{\delta q_{0}(X)^{2}\bar{\bg}_{0}\bar{\bg}_{0}^{\top}\}\Rightarrow E\{q_{0}(X)\bar{\bg}_{0}\bar{\bg}_{0}^{\top}\},\\
B_{24} & =E\left[\delta q_{0}(X)^{2}\left\{ \tau(X)-\tau_{0}\right\} \bar{\bg}_{0}\right]\Rightarrow E\left\{ q_{0}(X)\boldsymbol{\gamma}_{0}^{\top}\bar{\bg}_{0}\bar{\bg}_{0}\right\} ,\\
B_{33} & =E\{\delta q_{0}(X)^{2}V(A\vert X)\}\Rightarrow\pi_{A}(1-\pi_{A})E\left\{ q_{0}(X)\right\} ,\\
B_{34} & =E\{\delta q_{0}^{2}(X)\bar{\tau}(Y,A)(A-\pi_{A})\}=E\left[\delta q_{0}^{2}(X)\left\{ (1-\pi_{A})Y(1)+\pi_{A}Y(0)\right\} \right],\\
B_{44} & =E\{\delta q_{0}^{2}(X)\bar{\tau}(Y,A)^{2}\}=E\left[\delta q_{0}^{2}(X)\left\{ \frac{Y(1)^{2}}{\pi_{A}}+\frac{Y(0)^{2}}{1-\pi_{A}}-2\tau(X)\tau_{0}+\tau_{0}^{2}\right\} \right].
\end{align*}
We can express the asymptotic variance of $N^{-1/2}(\htau^{\mathrm{CW}}-\tau_{0})$
as $V^{\mathrm{CW}}=A_{41}B_{11}A_{41}^{\top}+A_{44}^{2}B_{44}$.

Under Assumption \ref{assump: outcome linear} and \ref{asump:logit linear},
we have $A_{41}B_{11}A_{41}^{\top}=E\left[\widetilde{\delta}d^{2}\left\{ \tau(X)-\tau_{0}\right\} ^{2}\right]$.
Therefore, $V^{\mathrm{CW}}$ can be simplified as 
\[
V^{\mathrm{CW}}=E\left[\widetilde{\delta}d^{2}\left\{ \tau(X)-\tau_{0}\right\} ^{2}\right]+E\left[\delta q_{0}^{2}(X)\left\{ \frac{Y(1)^{2}}{\pi_{A}}+\frac{Y(0)^{2}}{1-\pi_{A}}-2\tau(X)\tau_{0}+\tau_{0}^{2}\right\} \right].
\]

\subsection{Proof of Theorem~\ref{th:semi_bd-1}}

Let $Z=(X,A,Y,\delta,\tildelta)$ be a vector of random variables.
Assumptions \ref{a:consist} - \ref{a:pos} constitute the semiparametric
model. The semiparametric likelihood based on a single $Z$ is

\[
f(Z)=\left\{ f(X)f(A,Y\vert X,\delta=1)\pi_{\delta}(X)\right\} ^{\delta}\left\{ f(X)\right\} ^{\widetilde{\delta}},
\]
where $f(\cdot)$ is a density function for a continuous random variable
and is a probability mass function for a discrete random variable.

Assuming that $\delta\widetilde{\delta}=0$, the score function\citep{hahn1998role}
satisfies

\[
S(X,A,Y,\delta,\widetilde{\delta})=S(X,A,Y,\delta)+\widetilde{\delta}S(X).
\]

We first list four identities that are used in the following derivation
of the efficiency bound:
\begin{enumerate}
\item [{I1.}] For any function $h(X,A,\delta)$, we have $E\{h(X,A,\delta)S(Y\vert X,A,\delta)\}=0;$
\item [{I2.}] any function $h(X,A,\delta)$, we have $E[h(X,A,\delta)\{Y-E(Y\vert X,A,\delta)\}]=0;$
\item [{I3.}] any $h(X,A,Y)$, if $E\left\{ \widetilde{\delta}h(X,A,Y)\right\} =0$,
we have $E\{\widetilde{\delta}h(X,A,Y)S(X,A,Y,\delta)\}=0$
\item [{I4.}] any $h(X,A,Y)$, if $E\left\{ \delta h(X,A,Y)\right\} =0$,
we have $E\left\{ \delta h(X,A,Y)\widetilde{\delta}S(X)\right\} =0.$
\end{enumerate}
To derive the semiparametric efficiency score, we use the method of
parametric submodel \citep{Bickel1993}. Let $\left\{ f_{t}(Z):t\in\mathbb{R}\right\} $
be a regular parametric submodel which contains the truth at $t=0$,
i.e., $f_{t}(Z)\vert_{t=0}=f(Z).$

Note that $\tau(X)=E(Y\vert X,A=1,\delta=1)-E(Y\vert X,A=0,\delta=1)$
and $\tau_{0}=E\left\{ \widetilde{\delta}d\tau(X)\right\} $. Let
$\tau_{t}=E_{t}\left\{ \widetilde{\delta}d\tau_{t}(X)\right\} $ denote
the parameter $\tau$ evaluated with respect to the regular parametric
submodel $f_{t}(Z)$. Following \citet{Bickel1993}, the semiparametric
efficiency score $\phi=\phi(Z)$ is the pathwise derivative of the
target parameter in the sense that 
\[
\left.\frac{\partial}{\partial t}\tau_{t}\right\rvert _{t=0}=E\{\phi S(Z)\},
\]
where $S(Z)=\left.\partial\mathrm{log}f_{t}(Z)/\partial t\right\rvert _{t=0}$.
Toward this end, we express

\begin{equation}
\left.\frac{\partial}{\partial t}\tau_{t}\right\rvert _{t=0}=E\left\{ \widetilde{\delta}d\tau(X)S(X)\right\} +E\left.\left\{ \frac{\partial\tau_{t}(X)}{\partial t}\right\rvert _{t=0}\right\} .\label{eq:tau_t}
\end{equation}
For the first term in the right hand side of (\ref{eq:tau_t}), we
have

\begin{align}
E\left\{ \tildelta d\tau(X)S(X)\right\}  & =E\left[\tildelta d\left\{ \tau(X)-\tau_{0}\right\} S(X)\right]\nonumber \\
 & =E\left[\tildelta d\left\{ \tau(X)-\tau_{0}\right\} \left\{ S(X,A,Y,\delta)+\tildelta S(X)\right\} \right]\label{line4}\\
 & =E\left[\tildelta d\left\{ \tau(X)-\tau_{0}\right\} S(X,A,Y,\delta,\tildelta)\right],\nonumber 
\end{align}
where \eqref{line4} holds because of identity I3. To express further
the second term in the right hand side of (\ref{eq:tau_t}), we have
\begin{align*}
\left.\frac{\partial\tau_{t}(X)}{\partial t}\right\rvert _{t=0} & =\left.\left\{ \int y\frac{\partial}{\partial t}f_{t}(y\vert X,\delta=1,A=1)dy-\int y\frac{\partial}{\partial t}f_{t}(y\vert X,\delta=1,A=0)dy\right\} \right\rvert _{t=0}\\
 & =\left.\left\{ \int yS(y\vert X,A=1,\delta=1)f_{t}(y\vert X,A=1,\delta=1)dy\right\} \right\rvert _{t=0}\\
 & \qquad-\left.\left\{ \int yS(y\vert X,A=0,\delta=1)f_{t}(y\vert X,A=0,\delta=1)dy\right\} \right\rvert _{t=0}\\
 & =E\left\{ \frac{\delta AY}{\pi_{\delta}(X)\pi_{A}}S(Y\vert A,X,\delta)\vert X\right\} -E\left\{ \frac{\delta(1-A)Y}{\pi_{\delta}(X)(1-\pi_{A})}S(Y\vert A,X,\delta)\vert X\right\} \\
 & =E\left[\frac{\delta}{\pi_{\delta}(X)}\left\{ \frac{AY}{\pi_{A}}-\frac{(1-A)Y}{(1-\pi_{A})}\right\} S(Y\vert A,X,\delta)\vert X\right].
\end{align*}
Therefore, 
\begin{align}
E\left.\left\{ \frac{\partial\tau_{t}(X)}{\partial t}\right\rvert _{t=0}\right\}  & =E\left[\frac{\delta}{\pi_{\delta}(X)}\left\{ \frac{AY}{\pi_{A}}-\frac{(1-A)Y}{(1-\pi_{A})}\right\} S(Y\vert A,X,\delta)\right]\nonumber \\
 & =E\left(\frac{\delta}{\pi_{\delta}(X)}\left[\frac{A\{Y-\mu_{1,\delta}(X)\}}{\pi_{A}}-\frac{(1-A)\{Y-\mu_{0,\delta}(X)\}}{1-\pi_{A}}\right]S(Y\vert A,X,\delta)\right)\label{line1}\\
 & =E\left(\frac{\delta}{\pi_{\delta}(X)}\left[\frac{A\{Y-\mu_{1,\delta}(X)\}}{\pi_{A}}-\frac{(1-A)\{Y-\mu_{0,\delta}(X)\}}{1-\pi_{A}}\right]S(X,A,Y,\delta)\right)\label{line2}\\
 & =E\left(\frac{\delta}{\pi_{\delta}(X)}\left[\frac{A\{Y-\mu_{1,\delta}(X)\}}{\pi_{A}}-\frac{(1-A)\{Y-\mu_{0,\delta}(X)\}}{1-\pi_{A}}\right]S(X,A,Y,\delta,\widetilde{\delta})\right).\label{line3}
\end{align}
In the above derivation, \eqref{line1} follows by identity I1, \eqref{line2}
follows by identity I2, and \eqref{line3} follows by identities I3
and I4.

Substituting back to (\ref{eq:tau_t}), we have
\begin{small}
\[
\left.\frac{\partial}{\partial t}\tau_{t}\right\rvert _{t=0}=E\left\{ \left(\tildelta d\left\{ \tau(X)-\tau_{0}\right\} +\frac{\delta}{\pi_{\delta}(X)}\left[\frac{A\{Y-\mu_{1,\delta}(X)\}}{\pi_{A}}-\frac{(1-A)\{Y-\mu_{0,\delta}(x)\}}{1-\pi_{A}}\right]\right)S(X,A,Y,\delta,\widetilde{\delta})\right\} 
\]
\end{small}

Thus, the semiparametric efficiency score is 
\[
\phi=\widetilde{\delta}d\left\{ \tau(X)-\tau_{0}\right\} +\frac{\delta}{\pi_{\delta}(X)}\left[\frac{A\{Y-\mu_{1,\delta}(X)\}}{\pi_{A}}-\frac{(1-A)\{Y-\mu_{0,\delta}(X)\}}{(1-\pi_{A})}\right].
\]
It follows that the semiparametric efficiency bound is 
\[
E(\phi^{2})=E\left[\widetilde{\delta}d^{2}\left\{ \tau(X)-\tau_{0}\right\} ^{2}+\frac{\delta}{\pi_{\delta}^{2}(X)}\left\{ \frac{V\left\{ Y(1)|X,\delta\right\} }{\pi_{A}}+\frac{V\left\{ Y(0)|X,\delta\right\} }{1-\pi_{A}}\right\} \right].
\]

\subsection{Proof of Theorem \ref{th:consistency2}}

\subsubsection*{Proof of the double robustness of the ACW estimator}

Let $\boldsymbol{\theta}=(\bEta,\bbeta_{0},\bbeta_{1})$ denote the
vector of nuisance parameters. Note that $\htau^{\mathrm{ACW}}$ is
the solution to the estimating equation $N^{-1}\sum_{i=1}^{N}\phi(X_{i},A_{i},Y_{i},\delta_{i},\tildelta_{i};\tau,\widehat{\btheta})=0$,
where 
\begin{multline*}
\phi(X,A,Y,\delta,\widetilde{\delta};\tau,\btheta)=\frac{\delta}{\pi_{\delta}(X;\bEta)}\left[\frac{A\left\{ Y-\mu_{1,\delta}(X;\bbeta_{1})\right\} }{\pi_{A}}-\frac{(1-A)\left\{ Y-\mu_{0,\delta}(X;\bbeta_{0})\right\} }{1-\pi_{A}}\right]\\
+\widetilde{\delta}d\left\{ \mu_{1,1}(X;\bbeta_{1})-\mu_{0,1}(X;\bbeta_{0})\right\} -\tau.
\end{multline*}
Let $\boldsymbol{\theta}^{*}$ be the probability limits of $\widehat{\btheta}$.
It suffices to show that $E\left\{ \phi(X,A,Y,\delta,\widetilde{\delta};\tau_{0},\btheta^{*})\right\} =0$
if either $\pi(X;\bEta)$ or $\mu_{a,1}(X;\bbeta_{a})\,(a=0,1$)
is correctly specified. Under standard regularity conditions for M-estimators,
$\htau^{\mathrm{ACW}}$ is consistent for $\tau$. Use iterated expectation,
we can write

\begin{multline}
E\left\{ \phi(X,A,Y,\delta,\widetilde{\delta};\tau_{0},\btheta^{*})\right\} =E\left[\frac{\delta}{\pi_{\delta}(X;\bEta^{*})}\left\{ \frac{AY}{\pi_{A}}-\frac{(1-A)Y}{1-\pi_{A}}\right\} -\tau_{0}\right]+\\
E\left[\left\{ \widetilde{\delta}d-\frac{\delta}{\pi_{\delta}(X;\bEta^{*})}\right\} \left\{ \mu_{1,1}(X;\bbeta_{1}^{*})-\mu_{0,1}(X;\bbeta_{0}^{*})\right\} \right].\label{eq:Eofphi}
\end{multline}
The first term on the left-hand side of \eqref{eq:Eofphi} is $0$
either one of the $\pi(X;\bEta)$ or $\mu_{a,1}(X;\bbeta_{a})\,(a=0,1$)
is correctly specified, as shown in the proof of consistency in the
CW estimators. Now consider the second term on the left-hand side
of \eqref{eq:Eofphi}.

Firstly, if $\pi_{\delta}(X;\bEta)$ is correctly specified, we have
$\pi_{\delta}(X;\bEta^{*})=\pi_{\delta}(X)$. Take iterated expectation
conditional on $X$, we have the second term on the left-hand side
of \eqref{eq:Eofphi}

\begin{multline*}
E\left[\left\{ \widetilde{\delta}d-\frac{\delta}{\pi_{\delta}(X)}\right\} \left\{ \mu_{1,1}(X;\bbeta_{1}^{*})-\mu_{0,1}(X;\bbeta_{0}^{*})\right\} \right]\\
=E\left[\left\{ \mu_{1,1}(X;\bbeta_{1}^{*})-\mu_{0,1}(X;\bbeta_{0}^{*})\right\} E\left\{ \widetilde{\delta}d-\frac{\delta}{\pi_{\delta}(X)}\vert X\right\} \right]=0,
\end{multline*}
as $E\left\{ \widetilde{\delta}d-\delta/\pi_{\delta}(X)\vert X\right\} =0$.
Thus, \eqref{eq:Eofphi} equals to zero.

Secondly, if outcome model $\mu_{a,1}(X;\bbeta_{a})\,(a=0,1$)
is correctly specified, we have $\mu_{1,1}(X;\bbeta_{1}^{*})-\mu_{0,1}(X;\bbeta_{0}^{*})=\boldsymbol{\gamma}_{0}^{\top}\bg(X)$.
Then the second term on the left-hand side of \eqref{eq:Eofphi} satisfies

\[
E\left[\left\{ \widetilde{\delta}d-\frac{\delta}{\pi_{\delta}(X)}\right\} \left\{ \mu_{1,1}(X;\bbeta_{1}^{*})-\mu_{0,1}(X;\bbeta_{0}^{*})\right\} \right]=E\left[\boldsymbol{\gamma}_{0}^{\top}\left\{ \widetilde{\delta}d-\delta/\pi_{\delta}(X;\bEta^{*})\right\} \bg(X)\right]=0
\]
by the balancing constraint \eqref{eq:calibration constraints}. Thus,
\eqref{eq:Eofphi} equals to zero under this scenario as well. This
completes the proof of the double robustness of $\htau^{\mathrm{ACW}}$.

\subsection{Proof of Theorem \ref{th:consistency2} and Theorem \ref{th:semi_nuisance}\label{subsec:localEff}}

\subsubsection*{Proof of local efficiency}

Following the empirical process literature, let $\mathbb{P}_{N}$
denote the empirical measure. For a random variable $V$, $\mathbb{P}\{f(V)\}=\int f(v)d\mathbb{P}$
is the expectation of $f(V)$ under the true data-generating process.
Recall that $Z=(X,A,Y,\delta,\widetilde{\delta})$ , $\boldsymbol{\theta}=(\bEta,\bbeta_{0},\bbeta_{1})$,
$\boldsymbol{\theta}^{*}$ is the probability limits of $\widehat{\boldsymbol{\theta}}$
and $\boldsymbol{\theta}_{0}$ is the corresponding true parameter
value. Let

\begin{eqnarray*}
\psi(Z;\boldsymbol{\theta}) & = & \frac{\delta}{\pi_{\delta}(X;\bEta)}\left[\frac{A\left\{ Y-\mu_{1,1}(X;\bbeta_{1})\right\} }{\pi_{A}}-\frac{(1-A)\left\{ Y-\mu_{0,1}(X;\bbeta_{0})\right\} }{1-\pi_{A}}\right]\\
 &  & +\widetilde{\delta}d\left\{ \mu_{1,1}(X;\bbeta_{1})-\mu_{0,1}(X;\bbeta_{0})\right\} \\
 & = & \frac{\delta}{\pi_{\delta}(X;\bEta)}\frac{A\{Y-\mu_{1,1}(X;\bbeta_{1})\}}{\pi_{A}}+\tildelta d\mu_{1,1}(X;\bbeta_{1})\\
 &  & -\frac{\delta}{\pi_{\delta}(X;\bEta)}\frac{(1-A)\{Y-\mu_{0,1}(X;\bbeta_{0})\}}{1-\pi_{A}}-\tildelta d\mu_{0,1}(X;\bbeta_{0})\\
 & \eqqcolon & \psi_{1}(Z;\btheta)-\psi_{0}(Z;\btheta).
\end{eqnarray*}
Under the conditions specified in Theorem \ref{th:consistency2} or
the conditions specified in Theorem \ref{th:semi_nuisance}, and assume
that $\psi(Z;\boldsymbol{\theta})$ belongs to Donsker classes \citep{van1996weak,kennedy2016semiparametric},
$\mathbb{P}\psi_{1}(Z;\btheta^{*})=\mu_{1}$, $\mathbb{P}\psi_{0}(Z;\btheta^{*})=\mu_{0}$
and $\mathbb{P}\psi(Z;\btheta^{*})=\mu_{1}-\mu_{0}=\tau_{0}$. Thus,
\begin{align}
\htau^{\mathrm{ACW}}-\tau_{0} & =\mathbb{P}_{N}\psi(Z;\widehat{\btheta})-\mathbb{P}\psi(Z;\btheta^{*})\nonumber \\
 & =(\mathbb{P}_{N}-\mathbb{P})\psi(Z;\widehat{\boldsymbol{\theta}})+\mathbb{P}\{\psi(Z;\widehat{\boldsymbol{\theta}})-\psi(Z;\boldsymbol{\theta}^{*})\}\nonumber \\
 & =(\mathbb{P}_{N}-\mathbb{P})\psi(Z;\boldsymbol{\theta}^{*})+\mathbb{P}\{\psi(Z;\widehat{\boldsymbol{\theta}})-\psi(Z;\boldsymbol{\theta}^{*})\}+o_{p}(N^{-1/2}).\label{eq:tau_tau0}
\end{align}
We now show that 
\[
\mathbb{P}\{\psi(Z;\widehat{\boldsymbol{\theta}})-\psi(Z;\boldsymbol{\theta}^{*})\}=\mathbb{P}\{\psi_{1}(Z;\widehat{\boldsymbol{\theta}})-\psi_{1}(Z;\boldsymbol{\theta}^{*})\}-\mathbb{P}\{\psi_{0}(Z;\widehat{\boldsymbol{\theta}})-\psi_{0}(Z;\boldsymbol{\theta}^{*})\}
\]
is a small order term under conditions in Theorem \ref{th:consistency2}
or Theorem \ref{th:semi_nuisance}. We write

\begin{align*}
\mathbb{P}\{\psi_{1}(Z;\widehat{\boldsymbol{\theta}})-\psi_{1}(Z;\boldsymbol{\theta}^{*})\}= & \mathbb{P}\left[\frac{\delta}{\pi_{\delta}(X;\widehat{\bEta})}\frac{A\left\{ Y-\mu_{1,1}(X;\widehat{\bbeta}_{1})\right\} }{\pi_{A}}+\widetilde{\delta}d\mu_{1,1}(X;\widehat{\bbeta}_{1})-\mu_{1}\right]\\
= & \mathbb{P}\left[\left\{ \frac{\delta}{\pi_{\delta}(X;\widehat{\bEta})}-1\right\} \left\{ \mu_{1,1}(X)-\mu_{1,1}(X;\widehat{\bbeta}_{1})\right\} +\left(\widetilde{\delta}d-1\right)\mu_{1,1}(X;\widehat{\bbeta}_{1})\right]\\
= & \mathbb{P}\left[\left\{ \frac{\pi_{\delta}(X)-\pi_{\delta}(X;\widehat{\bEta})}{\pi_{\delta}(X;\widehat{\bEta})}\right\} \left\{ \mu_{1,1}(X)-\mu_{1,1}(X;\widehat{\bbeta}_{1})\right\} \right].
\end{align*}
Similarly, we have 
\[
\mathbb{P}\{\psi_{0}(Z;\widehat{\boldsymbol{\theta}})-\psi_{0}(Z;\boldsymbol{\theta}^{*})\}=\mathbb{P}\left[\left\{ \frac{\pi_{\delta}(X)-\pi_{\delta}(X;\widehat{\bEta})}{\pi_{\delta}(X;\widehat{\bEta})}\right\} \left\{ \mu_{0,1}(X)-\mu_{0,1}(X;\widehat{\bbeta}_{0})\right\} \right].
\]
Therefore, by Cauchy-Schwarz inequality and the positivity of $\pi_{\delta}(X;\widehat{\bEta})$,
$\vert\mathbb{P}\{\psi(Z;\widehat{\boldsymbol{\theta}})-\psi(Z;\boldsymbol{\theta}^{*})\}\vert$
is bounded above by 
\begin{equation}
\lVert\pi_{\delta}(X)-\pi_{\delta}(X;\widehat{\bEta})\rVert\sum_{a\in\{0,1\}}\rVert\mu_{a,1}(X)-\mu_{a,1}(X;\widehat{\bbeta}_{a})\rVert.\label{eq:upperBound}
\end{equation}
Under the conditions in Theorem \ref{th:consistency2}, if $\pi_{\delta}(X;\bEta)$
is a correctly specified parametric model for $\pi_{\delta}(X)$,
then $\lVert\pi_{\delta}(X)-\pi_{\delta}(X;\widehat{\bEta})\rVert=O_{p}(n^{-1/2})$;
and if $\mu_{a,1}(X;\bbeta_{a})$ is a correctly specified
parametric model for $\mu_{a,1}(X)$, then $\rVert\mu_{a,1}(X)-\mu_{a,1}(X;\widehat{\bbeta}_{a})\rVert=O_{p}(n^{-1/2})$.
Therefore, the product \eqref{eq:upperBound} is $O_{p}(n^{-1})$,
which makes $\mathbb{P}\{\psi(Z;\widehat{\boldsymbol{\theta}})-\psi(Z;\boldsymbol{\theta}^{*})\}$
in \eqref{eq:tau_tau0} asymptotically negligible. Under the conditions
in Theorem \ref{th:semi_nuisance}, the product \eqref{eq:upperBound}
is $o_{p}(n^{-1/2})$ and therefore the term $\mathbb{P}\{\psi(Z;\widehat{\boldsymbol{\theta}})-\psi(Z;\boldsymbol{\theta}^{*})\}$
in \eqref{eq:tau_tau0} is asymptotically negligible. The result follows.

\subsection{Proof of Lemma \ref{lm:varcomparison}} \label{subsec:varcomparison}
Since $\pi_{\delta}(X^{O})=\pi_{\delta}(X^{C})$ and $V\{Y(a)\mid X^{O}\}\leq V\{Y(a)\mid X^{C}\}$ for $a = 0,1$,
\begin{align*}
    V_{1}^{O} &= E\left[\frac{1}{\pi_{\delta}(X^{O})}\left\{ \frac{V\{Y(1)\mid X^{O}\}}{\pi_{A}}+\frac{V\{Y(0)\mid X^{O}\}}{1-\pi_{A}}\right\} \right] \\
                & = E\left[\frac{1}{\pi_{\delta}(X^{C})}\left\{ \frac{V\{Y(1)\mid X^{O}\}}{\pi_{A}}+\frac{V\{Y(0)\mid X^{O}\}}{1-\pi_{A}}\right\} \right] \\
                & \le E\left[\frac{1}{\pi_{\delta}(X^{C})}\left\{ \frac{V\{Y(1)\mid X^{C}\}}{\pi_{A}}+\frac{V\{Y(0)\mid X^{C}\}}{1-\pi_{A}}\right\} \right] = V_{1}^{C},
\end{align*}
which proves the first part of inequalities.

For the second part, since $V\{Y(a)\mid X^{C},X^{I}\} =  V\{Y(a)\mid X^{C}\}$ for $a = 0,1$, 
\begin{align*}
    V_{1}^{C+I} & = E\left[\frac{1}{\pi_{\delta}(X^{C},X^{I})}\left\{ \frac{V\{Y(1)\mid X^{C},X^{I}\}}{\pi_{A}}+\frac{V\{Y(0)\mid X^{C},X^{I}\}}{1-\pi_{A}}\right\} \right]\\
    & =E\left[E\left\{ \frac{1}{\pi_{\delta}(X^{C},X^{I})}\mid X^{C}\right\} \left\{ \frac{V\{Y(1)\mid X^{C}\}}{\pi_{A}}+\frac{V\{Y(0)\mid X^{C}\}}{1-\pi_{A}}\right\} \right]\\
    &\geq E\left[\frac{1}{E\{\pi_{\delta}(X^{C},X^{I})\mid X^{C}\}}\left\{ \frac{V\{Y(1)\mid X^{C}\}}{\pi_{A}}+\frac{V\{Y(0)\mid X^{C}\}}{1-\pi_{A}}\right\} \right]\\
    &= E\left[\frac{1}{\pi_{\delta}(X^{C})}\left\{ \frac{V\{Y(1)\mid X^{C}\}}{\pi_{A}}+\frac{V\{Y(0)\mid X^{C}\}}{1-\pi_{A}}\right\} \right]=V_{1}^{C},
\end{align*}
where the second and third lines are from the double expectation and Jensen's inequality, respectively.

\section{Conditions for the Sieves Estimator \label{s:Conditions}}
Following \citet{hirano2003efficient}, we assume the following regularity
conditions on the data generating process and the nuisance functions.

\begin{condition}[Distribution of X]
Let $\mathcal{X}\subseteq\mathbb{R}^{p}$ be the support of $X$.
Assume that $\mathcal{X}$ is a Cartesian product of compact intervals,
i.e., $\mathcal{X}=\prod_{j=1}^{p}\left[l_{j},u_{j}\right]$, $l_{j},u_{j}\in\mathbb{R}$.
The density of $X$, $f(X)$, is bounded above and below away from
$0$ on $\mathcal{X}$.
\end{condition}

\begin{condition}[Basis functions]There exist constant $l$ and
$u$ such that 
$$
l\le\rho_{\mathrm{min}}\{\bg(X)^{\top}\bg(X)\}\le\rho_{\mathrm{max}}\{\bg(X)^{\top}\bg(X)\}\le u,
$$
almost surely where $\rho_{\mathrm{min}}$ and $\rho_{\mathrm{\mathrm{max}}}$
denote the minimum and maximum eigenvalues of a matrix.

\end{condition}

\begin{condition}[Potential outcomes]

The second moment of the potential outcomes are finite. i.e., $E\left\{ Y(a)^{2}\right\} <\infty$,
for $a=0,1$.\end{condition}

\begin{condition}[Smoothness]

The log sampling score function $\log\pi_{\delta}(x)$ is $s_{\delta}$-times
continuously differentiable and the outcome mean function $\mu_{a}(x)$
is $s_{\mu_{a}}$-times continuously differentiable, $\forall x\in\mathcal{X},\;a=0,1$;
The sieves estimators of $\log\pi_{\delta}(x)$ and $\mu_{a}(x)$
use a power series; the smoothness condition is $s>4p$, for $s=s_{\delta}$
and $s=s_{\mu_{a}}\;(a=0,1)$, respectively.

\end{condition}

The constraint $s>4p$ is required such that $\nu$ exists in the
following condition for the number of basis functions.

\begin{condition}[Number of basis function]\label{cond:K}

The number of basis functions $K$ satisfies $K=O(n^{\nu})$, where
$p/(2s-4p)<\nu<1/4$.

\end{condition}

Under the above conditions, the bias of the sieves approximations
are $O_{p}(K^{1-s/(2p)})=o_{p}(n^{-1/4})$. Moreover, because $K^{4}=o(n)$,
the variances of the sieves approximations are $O_{p}(K/n)=o_{p}(n^{-1/2})$.

To present regularity conditions for the penalization approach to
choosing $K,$ we introduce more notation. Let the support of model
parameters be 
\[
\M_{\delta}=\{1\leq j\leq\widetilde{K}:\bEta_{j}^{*}\neq0\},\ \ \M_{a}=\{1\leq j\leq\widetilde{K}:\bbeta_{a,j}^{*}\neq0\},\ (a=0,1).
\]
Define $K_{\bEta}=||\bEta^{*}||_{0}$, $K_{a}=||\bbeta_{a}^{*}||_{0}$
$(a=0,1)$, $K=\max(K_{\bEta},K_{0},K_{1})$, and $\xi_{\min}=\min(\xi,\xi_{0},\xi_{1})$.
Let $C$, $C_{1}$ and $C_{2}$ be generic constants. For any $\J\subseteq\{1,\ldots,\widetilde{K}\}$
and any vector $\bEta\in\R^{\widetilde{K}}$, let $\bEta_{\J}$ be
the sub-vector of $\bEta$ formed by elements of $\bEta$ whose indexes
are in $\J$. Let $\J^{c}$ be the complement of $\J$.

\begin{assumption}\label{assump:regularity } The following regularity
conditions hold.
\begin{description}
\item [{(A1)}] The parameter $(\bEta^{\T},\bbeta_{0}^{\T},\bbeta_{1}^{\T})^{\T}$
belongs to a compact subset in $\R^{3\widetilde{K}}$, and $(\bEta^{*\T},\bbeta_{0}^{*\T},\bbeta_{1}^{*\T})^{\T}$
lies in the interior of the compact subset.
\item [{(A2)}] Let $\epsilon_{\delta,i}(\bEta)=\delta_{i}-\exp\{\bEta^{\top}\bg(X_{i})\}$.
There exists a constant $C$ such that $E\{|\epsilon_{\delta,i}(\bEta^{*})|^{2+\delta}\}\leq C$
for all $i$ and some $\delta>0$. There exist constants $C_{1}$
and $C_{2}$ such that $E[\exp\{C_{1}|\epsilon_{\delta,i}(\bEta^{*})|\}\mid X_{i}]\leq C_{2}$
for all $i$.
\item [{(A3)}] $\exp^{(1)}\{\bEta^{\top}\bg(X_{i})\}$, $\exp^{(2)}\{\bEta^{\top}\bg(X_{i})\}$,
and $\exp^{(3)}\{\bEta^{\top}\bg(X_{i})\}$ (which denote the first,
second and third derivative of $\exp\{\bEta^{\top}\bg(X_{i})\}$,
respectively) are uniformly bounded away from $\infty$ on $\N_{\delta,\tau}=\{\bEta\in\R^{\widetilde{K}}:||\bEta_{\M_{\delta}}-\bEta_{\M_{\delta}}^{*}||\leq\tau\sqrt{K/n},\bEta_{\M_{\delta}^{c}}=0\}$
for some $\tau>0$.
\item [{(A4)}] For $a=0,1,$ let $\epsilon_{a,i}(\bbeta_{a})=Y_{i}(a)-m_{a}\{\bbeta_{a}^{\top}\bg(X_{i})\}$.
There exists a constant $C$ such that $E\{|\epsilon_{a,i}(\bbeta_{a}^{*})|^{2+\delta}\}\leq C$
for all $i$ and some $\delta>0$. There exist constants $C_{1}$
and $C_{2}$ such that $E[\exp\{C_{1}|\epsilon_{a,i}(\bbeta_{a}^{*})|\}\mid X_{i}]\leq C_{2}$
for all $i$.
\item [{(A5)}] For $a=0,1,$ $m_{a}^{(1)}\{\bbeta_{a}^{\top}\bg(X)\}$,
$m_{a}^{(2)}\{\bbeta_{a}^{\top}\bg(X)\}$, and $m_{a}^{(3)}\{\bbeta_{a}^{\top}\bg(X)\}$
(which denote the first, second and third derivative of $m_{a}\{\bbeta_{a}^{\top}\bg(X)\}$,
respectively) are uniformly bounded away from $\infty$ on $\N_{a,\tau}=\{\bbeta_{a}\in\R^{\widetilde{K}}:||\bbeta_{a,\M_{a}}-\bbeta_{a,\M_{a}}^{*}||\leq\tau\sqrt{K/n},\bbeta_{a,\M_{a}^{c}}=0\}$
for some $\tau>0$.
\item [{(A6)}] $\min_{j\in\M_{\delta}}|\eta_{j}^{*}|/\xi\rightarrow\infty$
and $\min_{j\in\M_{a}}|\beta_{a,j}^{*}|/\xi_{a}\rightarrow\infty,$
($a=0,1$) as $n\rightarrow\infty$.
\item [{(A7)}] $K=o(n^{1/3})$, $\xi_{\min}\rightarrow0$, $(\log n)^{2}=o(n\xi_{\min}^{2})$,
$\log(\widetilde{K})=o\left\{ n\xi_{\min}^{2}/\left(\log n\right){}^{2}\right\} $,
$\widetilde{K}K^{4}(\log n)^{6}=o(n^{3}\xi_{\min}^{2})$, $\widetilde{K}K^{4}(\log n)^{8}=o(n^{4}\xi_{\min}^{4})$,
as $n\rightarrow\infty$.
\end{description}
\end{assumption}

These assumptions are typical in the penalization literature.\textcolor{black}{{}
Assumptions (A2) and }(A4) hold for Gaussian distribution, sub-Gaussian
distribution, and so on. Assumptions (A3) and (A5) hold for common
models. Assumption (A7) specifies the restrictions on the initial
number of sieves functions $\widetilde{K}$ and the maximum dimension
of the true nonzero coefficients $K$. To gain insight, when the true
model size $K=O(n^{\nu})$, where $\nu$ satisfies Condition \ref{cond:K},
then $\widetilde{K}=O(n^{\widetilde{\nu}})$ with $\widetilde{\nu}<2-4\nu$
meets the (A7) requirement.

\section{Simulation Study \label{s:Simulation}}

\subsection{Simulation Details \label{s:simulation details}}
In this section, we evaluate the finite sample performances of the
proposed estimators via a set of simulation experiments. 
Covariate
$X\in\R^{5}$ is generated from $X_{j}\sim\N(1,1)$ for each $j=1,\dots,5$.
We generate potential outcome according to 

\begin{equation*}
    Y(a)\vert X = -100+27.4aX_{3}+13.7X_{4}+10aX_{4}+13.7X_{5} - 10aX_{5}+\epsilon,
\end{equation*}

where $\log\epsilon\sim\mathcal{N}(0,0.25)$ for $a=0,1$. Under this
setting, the true ATE is $\tau_{0}=27.4$. From the hypothetical
RCT eligible population, we generate the indicator of 
trial participation according to $\delta\mid X\sim\mathrm{Bernoulli}\{\pi_{\delta}(X)\},$
where $\log\{\pi_{\delta}(X)\}=-7.7+2X_{1} + 0.3X_{2}-0.4X_{3}.$ 
Note that $X_3, X_4, X_5$ are outcome predictors, and $X_1$ is an IV that is highly predictive of trial participation.

By this design, there are approximately $n=440$ subjects in the RCT
sample. The treatment assignment in the RCT sample is $A\sim\mathrm{Bernoulli}(0.5)$.
From the observational study population, we select a random sample
of size $m=2000$ to form an observational sample. For subjects
in the observational sample, the treatment assignment is generated
by $A\mid X\sim\mathrm{Bernoulli}\{e_{A}(X)\},$ where $\logit\{e_{A}(X)\}=-X_{1}+0.4X_{2}-0.25X_{3}-0.1X_{4}+0.1X_{5}.$
The actual observed outcome $Y$  is  $Y=AY(1)+(1-A)Y(0)$.

To study the impact of model misspecification, adapting from \citet{kang2007demystifying},
we define a nonlinear transformation of $X$ to be 
\begin{equation*}
    X^{*}= \left[\mathrm{exp}\left(X_{1}/10\right),(X_{3}+X_{5}+20)^{2},X_{2}/\{ 2+0.5\exp(X_{4})\},
    (X_{1}+X_{4}+20)^{2}, 0.5X_{2}X_{3}+X_{5} \right]^\top,
\end{equation*}
and further scale and center $X^{*}$ such that $E(X_{j}^{*})=1$
and $V(X_{j}^{*})=1$, for $j=1,\dots,5$. Throughout, we use $X$
for fitting models. We assume $X^{*}$ to be unobserved, but it can be
used in the true generative models, in which cases the fitted models
are misspecified.

To demonstrate the double robustness of the ACW estimator against
parametric model misspecification, we consider $\bg_{1}(X)=\left(X_{1},X_{2},X_{3},X_{4},X_{5}\right)^{\top}$
as the calibration variables in all four scenarios. Moreover, we consider the
 ACW estimator using sieves estimation, extending 
$\bg_{1}(X)$ to include further all two-way interaction terms and
quadratic terms, i.e. $\bg_{2}(X)=(X_{1},\dots,X_{5},X_{1}X_{2},\dots,X_{4}X_{5},X_{1}^{2},\dots,X_{5}^{2})^{\top}$.
We compare the proposed outcome-prioritized sieve method, denoted (S$^{O}$), with the less efficient method that constructs the sieve basis for $\pi_{\delta}(X)$ by power series of $X$ and solves penalized estimating equation, denoted (S).

Note that for ACW estimators, the outcome mean functions can be estimated by using either trial sample $(A,Y)$ only or both RCT and observational sample. We denote the former by ACW-t to indicate that only trial data is used and the latter by ACW-b to emphasize the use of both data sources.

We compare the proposed CW and ACW estimators with other ATE estimators in Table \ref{table:simulation} in the main paper.
We use bootstrap variance estimation for all estimators with $B=50$.
All simulations are based on 1000 Monte Carlo replications; Table~\ref{table:summary1_port} summarizes
the results. 

\begin{table}
\centering
\footnotesize
\caption{Simulation results under four scenarios. Bias is the empirical bias of point estimates; ESE is the empirical standard error of estimates; MSE is the mean squared error of estimates; RSE is the relative bias (\%) of bootstrap standard error estimates; CP is the empirical coverage probability of the 95\% confidence intervals.}
\label{table:summary1_port} %
\rowcolors{5}{white}{Gray}
    \resizebox{\textwidth}{!}{
\begin{tabular}{ccccccccccc}
\toprule 
& \multicolumn{5}{c}{\textbf{Scenario 1: O:C/S:C}} & \multicolumn{5}{c}{\textbf{Scenario 2: O:C/S:W}} \tabularnewline \cmidrule(lr){2-6} \cmidrule(lr){7-11}
 & \textbf{BIAS} & \textbf{ESE} &  \textbf{MSE} & \textbf{RSE(\%)} & \textbf{CP(\%)} & \textbf{BIAS} & \textbf{ESE} &  \textbf{MSE} & \textbf{RSE(\%)} & \textbf{CP(\%)} \tabularnewline
\midrule
Naive &-9.62 & 2.77 & 100.23 & -0.3 & 6.7 & 3.77 & 2.54 & 20.69 & -0.3 & 65.1 \tabularnewline
IPSW & -2.05 & 9.66 & 97.44 & -24.5 & 85.9 & 1.91 & 11.50 & 135.81 & -30.3 & 85.2 \tabularnewline
AIPSW & 0.03 & 0.67 & 0.45 & 3.0 & 95.5 & 0.03 & 0.70 & 0.49 & -1.1 & 94.8 \tabularnewline
AIPSW(S) & 0.03 & 0.67 & 0.45 & 2.9 & 95.6 & 0.03 & 0.70 & 0.49 & -1.2 & 94.8 \tabularnewline
SBW &  0.51 & 9.54 & 91.26 & 4.9 & 94.9 & 0.21 & 10.56 & 111.49 & -1.8 & 93.9 \tabularnewline
CW & 0.56 & 11.25 & 126.70 & -10.1 & 90.3 & 0.21 & 12.48 & 155.69 & -15.3 & 88.4 \tabularnewline
ACW-t &0.03 & 0.68 & 0.46 & 2.1 & 95.2 & 0.03 & 0.70 & 0.49 & -1.0 & 94.6 \tabularnewline
ACW-t(S) & 0.02 & 0.71 & 0.50 & 2.6 & 95.8 & 0.03 & 0.72 & 0.52 & 0.4 & 95.0 \tabularnewline
ACW-t$\left(\mbox{S}^{O}\right)$ & 0.03 & 0.67 & 0.45 & 3.2 & 95.7 & 0.03 & 0.70 & 0.48 & -1.0 & 94.3 \tabularnewline
ACW-b & 0.03 & 0.68 & 0.46 & 2.4 & 95.8 & 0.03 & 0.70 & 0.49 & -0.6 & 94.5 \tabularnewline
ACW-b(S) & 0.03 & 0.70 & 0.49 & 3.8 & 96.0 & 0.03 & 0.72 & 0.52 & 0.1 & 95.9 \tabularnewline
ACW-b$\left(\mbox{S}^{O}\right)$ & 0.03 & 0.66 & 0.44 & 3.0 & 95.2 & 0.03 & 0.69 & 0.48 & -1.5 & 95.2 \tabularnewline
\midrule
& \multicolumn{5}{c}{\textbf{Scenario 3: O:W/S:C}} & \multicolumn{5}{c}{\textbf{Scenario 4: O:W/S:W}} \tabularnewline \cmidrule(lr){2-6} \cmidrule(lr){7-11}
 \rowcolor{white} & \textbf{BIAS} & \textbf{ESE} &  \textbf{MSE} & \textbf{RSE(\%)} & \textbf{CP(\%)} & \textbf{BIAS} & \textbf{ESE} &  \textbf{MSE} & \textbf{RSE(\%)} & \textbf{CP(\%)} \tabularnewline
\midrule
Naive & 20.53 & 2.52 & 427.79 & -0.8 & 0.0 & 5.88 & 2.51 & 40.92 & -1.7 & 35.8 \tabularnewline
IPSW & 6.08 & 11.44 & 167.69 & -30.6 & 70.6 & -0.07 & 12.96 & 167.77 & -33.3 & 86.0 \tabularnewline
AIPSW & -1.84 & 3.23 & 13.82 & -27.2 & 75.0 & -2.45 & 3.98 & 21.87 & -35.1 & 69.1 \tabularnewline
AIPSW(S) & -0.10 & 0.84 & 0.71 & -12.1 & 90.5 & -0.07 & 0.82 & 0.68 & -15.0 & 90.8 \tabularnewline
SBW & 0.45 & 8.74 & 76.48 & 0.7 & 93.8 & -1.21 & 9.97 & 100.69 & -1.8 & 92.5 \tabularnewline
CW & 0.87 & 11.22 & 126.58 & -16.4 & 87.4 & -1.05 & 12.48 & 156.75 & -17.9 & 85.4 \tabularnewline
ACW-t & -0.15 & 3.59 & 12.86 & -14.9 & 90.2 & -1.40 & 3.98 & 17.77 & -17.2 & 86.6 \tabularnewline
ACW-t(S) & 0.05 & 1.44 & 2.07 & 21.1 & 96.5 & 0.05 & 1.35 & 1.84 & 2.7 & 95.1 \tabularnewline
ACW-t$\left(\mbox{S}^{O}\right)$ & -0.04 & 0.88 & 0.77 & -0.8 & 94.1 & -0.02 & 0.85 & 0.72 & -5.0 & 93.4 \tabularnewline
ACW-b & 0.14 & 3.41 & 11.62 & -15.6 & 91.4 & -1.18 & 3.65 & 14.68 & -19.5 & 87.5 \tabularnewline
ACW-b(S) & 0.15 & 2.37 & 5.64 & 3.6 & 96.6 & 0.18 & 2.17 & 4.74 & -1.8 & 97.0 \tabularnewline
ACW-b$\left(\mbox{S}^{O}\right)$ & 0.01 & 0.74 & 0.54 & 1.9 & 95.9 & 0.01 & 0.73 & 0.53 & 0.6 & 94.5 \tabularnewline
\bottomrule 
\end{tabular}
}
\end{table}

\newpage

\subsection{Additional Simulation: Three-way Calibration\label{s:three way}}
In this subsection, we present simulation results of the CW and ACW estimators that achieve the three-way balance between the treated RCT, the control RCT, and the observational sample, similar to \citet{chan2016globally}. We use the same simulation setting that is described in Section \ref{s:simulation details}. Table \ref{table:summary_3way} and Figure \ref{fig:compare3way} summarise the simulation results based on 1000 Monte Carlo replications. It can be seen that three-way calibration is comparable under Scenario 1 and Scenario 2 but more unstable under Scenario 3 and Scenario 4 than the two-way calibration in general.

\begin{table}[ht]
\centering
\footnotesize
\caption{Simulation results of the three-way calibration. Bias is the empirical bias of point estimates; ESE is the empirical standard error of estimates; MSE is the mean squared error of estimates; RSE is the relative bias (\%) of bootstrap standard error estimates; CP is the empirical coverage probability of the 95\% confidence intervals.}
\label{table:summary_3way} %
    \resizebox{\textwidth}{!}{
\begin{tabular}{ccccccccccc}
\toprule 
& \multicolumn{5}{c}{\textbf{Scenario 1: O:C/S:C}} & \multicolumn{5}{c}{\textbf{Scenario 2: O:C/S:W}} \tabularnewline \cmidrule(lr){2-6} \cmidrule(lr){7-11}
 & \textbf{BIAS} & \textbf{ESE} &  \textbf{MSE} & \textbf{RSE(\%)} & \textbf{CP(\%)} & \textbf{BIAS} & \textbf{ESE} &  \textbf{MSE} & \textbf{RSE(\%)} & \textbf{CP(\%)} \tabularnewline
\midrule
CW & 0.018 & 0.703 & 0.493 & -3.6 & 92.4 & 0.020 & 0.693 & 0.480 & -1.3 & 93.7\tabularnewline
\rowcolor{Gray} ACW-t & 0.017 & 0.743 & 0.552 & -6.7 & 91.9 & 0.020 & 0.723 & 0.522 & -3.2 & 92.7\tabularnewline
ACW-t(S) & 0.024 & 0.796 & 0.634 & -2.3 & 94.1 & 0.021 & 0.760 & 0.578 & 2.7 & 95.9\tabularnewline
\rowcolor{Gray} ACW-t$\left(\mbox{S}^{O}\right)$ & 0.015 & 0.695 & 0.483 & -1.2 & 94.1 & 0.020 & 0.689 & 0.474 & 0.6 & 95.1\tabularnewline
ACW-b & 0.016 & 0.743 & 0.552 & -6.0 & 91.8 & 0.020 & 0.722 & 0.522 & -3.9 & 91.8\tabularnewline
\rowcolor{Gray} ACW-b(S) & 0.018 & 0.785 & 0.615 & -0.1 & 94.4 & 0.023 & 0.770 & 0.593 & 2.2 & 96.0\tabularnewline
ACW-b$\left(\mbox{S}^{O}\right)$ & 0.019 & 0.687 & 0.472 & -0.1 & 94.7 & 0.019 & 0.683 & 0.466 & 1.0 & 95.4\tabularnewline
\midrule
& \multicolumn{5}{c}{\textbf{Scenario 3: O:W/S:C}} & \multicolumn{5}{c}{\textbf{Scenario 4: O:W/S:W}} \tabularnewline \cmidrule(lr){2-6} \cmidrule(lr){7-11}
 & \textbf{BIAS} & \textbf{ESE} &  \textbf{MSE} & \textbf{RSE(\%)} & \textbf{CP(\%)} & \textbf{BIAS} & \textbf{ESE} &  \textbf{MSE} & \textbf{RSE(\%)} & \textbf{CP(\%)} \tabularnewline
\midrule
CW & 0.042 & 3.293 & 10.837 & -31.4 & 81.1 & -1.333 & 3.664 & 15.188 & -37.5 & 75.8\tabularnewline
\rowcolor{Gray} ACW-t & 7.105 & 6.466 & 92.252 & -31.5 & 54.5 & 3.282 & 7.284 & 63.771 & -38.2 & 69.4\tabularnewline
ACW-t(S) & 0.221 & 2.658 & 7.108 & -6.5 & 96.6 & 0.292 & 2.273 & 5.247 & -8.4 & 95.8\tabularnewline
\rowcolor{Gray} ACW-t$\left(\mbox{S}^{O}\right)$ & 0.225 & 1.277 & 1.679 & -11.3 & 93.6 & 0.193 & 1.109 & 1.267 & -9.4 & 93.3\tabularnewline
ACW-b & 2.886 & 6.631 & 52.251 & -32.5 & 73.7 & 0.165 & 7.314 & 53.469 & -37.3 & 77.5\tabularnewline
\rowcolor{Gray} ACW-b(S) & 0.128 & 2.883 & 8.320 & 33.8 & 97.4 & 0.232 & 2.819 & 7.994 & 22.0 & 96.9\tabularnewline
ACW-b$\left(\mbox{S}^{O}\right)$ & 0.010 & 1.154 & 1.331 & -3.4 & 94.5 & 0.010 & 1.223 & 1.494 & -6.5 & 94.2\tabularnewline
\bottomrule 
\end{tabular}
}
\end{table}

\begin{figure}[h]
\centerline{
\includegraphics[width=6in]{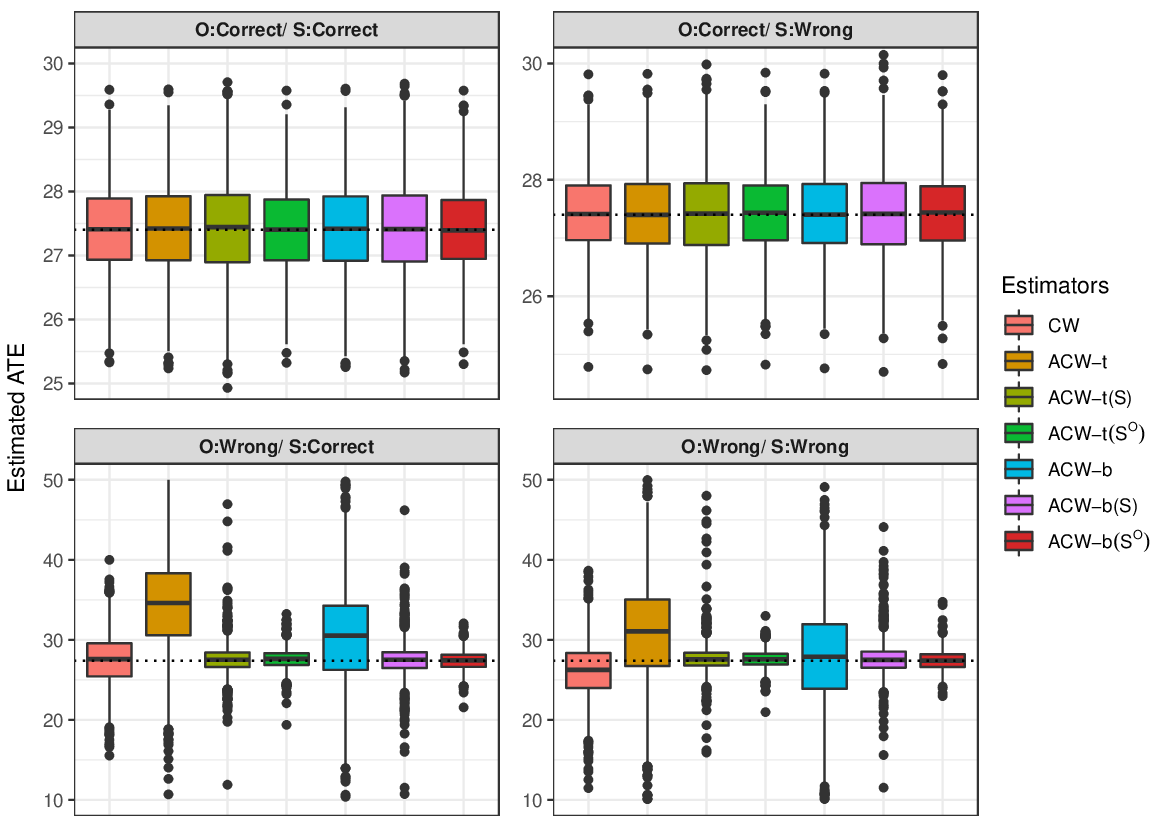}}
\caption{Boxplot of the CW and ACW estimators with three-way calibration between the treated RCT, the control RCT, and the observational sample. We removed a few outliers for visualization.\label{fig:compare3way}}
\end{figure}

\newpage

\subsection{Additional Simulation: Increased Observational Sample Size}
We consider increasing the observational sample size to $m = 15,000$ while keeping the RCT sample size and other simulation settings the same as those in Section \ref{s:simulation details}. This resembles the data structure of CALGB 9633 and NCDB described in Section \ref{sec:Real-data-application}. Table \ref{table:summary1_large} and Figure \ref{fig:comparelarge} summarize the results based on 500 Monte Carlo replications. It can be seen that when the sample sizes are larger, the advantages of our proposed estimators become more obvious. Especially in Scenario 3 and Scenario 4 where outcome models are not correct, the proposed ACW-t$\left(\mbox{S}^{O}\right)$ and ACW-b$\left(\mbox{S}^{O}\right)$ estimators outperform other doubly robust estimators in terms of robustness and efficiency.

\begin{table}[ht]
\centering
\footnotesize
\caption{Simulation results when the observational sample size is 15,000. Bias is the empirical bias of point estimates; ESE is the empirical standard error of estimates; MSE is the mean squared error of estimates; RSE is the relative bias (\%) of bootstrap standard error estimates; CP is the empirical coverage probability of the 95\% confidence intervals.}
\label{table:summary1_large} %
\rowcolors{5}{white}{Gray}
    \resizebox{\textwidth}{!}{
\begin{tabular}{ccccccccccc}
\toprule 
& \multicolumn{5}{c}{\textbf{Scenario 1: O:C/S:C}} & \multicolumn{5}{c}{\textbf{Scenario 2: O:C/S:W}} \tabularnewline \cmidrule(lr){2-6} \cmidrule(lr){7-11}
 & \textbf{BIAS} & \textbf{ESE} &  \textbf{MSE} & \textbf{RSE(\%)} & \textbf{CP(\%)} & \textbf{BIAS} & \textbf{ESE} &  \textbf{MSE} & \textbf{RSE(\%)} & \textbf{CP(\%)} \tabularnewline
\midrule
Naive & -9.717 & 2.876 & 102.683 & -3.5 & 8.0 & 3.709 & 2.534 & 20.161 & 0.1 & 68.6\tabularnewline
IPSW & -0.931 & 10.406 & 108.927 & -22.2 & 89.2 & 1.035 & 13.110 & 172.598 & -30.8 & 84.2\tabularnewline
AIPSW & -0.006 & 0.263 & 0.069 & -0.6 & 95.2 & 0.011 & 0.256 & 0.065 & 3.3 & 93.8\tabularnewline
AIPSW(S) & -0.006 & 0.267 & 0.071 & -1.6 & 94.6 & 0.012 & 0.265 & 0.070 & 2.1 & 94.8\tabularnewline
SBW & 0.368 & 8.615 & 74.210 & 14.2 & 95.6 & 0.222 & 10.318 & 106.305 & -2.6 & 94.0\tabularnewline
CW & 0.776 & 10.460 & 109.792 & -5.6 & 90.2 & -0.054 & 12.271 & 150.281 & -14.8 & 88.4\tabularnewline
ACW-t & -0.005 & 0.272 & 0.074 & 0.3 & 95.0 & 0.011 & 0.263 & 0.069 & 4.4 & 95.8\tabularnewline
ACW-t(S) & 0.004 & 0.311 & 0.096 & 9.0 & 95.6 & 0.009 & 0.327 & 0.107 & 3.3 & 95.2\tabularnewline
ACW-t$\left(\mbox{S}^{O}\right)$ & -0.006 & 0.246 & 0.061 & 4.8 & 95.8 & 0.005 & 0.236 & 0.056 & 7.4 & 96.4\tabularnewline
ACW-b & -0.005 & 0.273 & 0.075 & 0.0 & 94.8 & 0.010 & 0.263 & 0.069 & 4.3 & 95.0\tabularnewline
ACW-b(S) & 0.000 & 0.316 & 0.100 & 9.1 & 96.4 & 0.019 & 0.316 & 0.100 & 9.1 & 96.4\tabularnewline
ACW-b$\left(\mbox{S}^{O}\right)$ & 0.001 & 0.245 & 0.060 & 3.4 & 94.8 & 0.004 & 0.236 & 0.056 & 6.5 & 96.6\tabularnewline
\midrule
& \multicolumn{5}{c}{\textbf{Scenario 3: O:W/S:C}} & \multicolumn{5}{c}{\textbf{Scenario 4: O:W/S:W}} \tabularnewline \cmidrule(lr){2-6} \cmidrule(lr){7-11}
 \rowcolor{white} & \textbf{BIAS} & \textbf{ESE} &  \textbf{MSE} & \textbf{RSE(\%)} & \textbf{CP(\%)} & \textbf{BIAS} & \textbf{ESE} &  \textbf{MSE} & \textbf{RSE(\%)} & \textbf{CP(\%)} \tabularnewline
\midrule
Naive & 20.481 & 2.613 & 426.273 & -3.9 & 0.0 & 5.924 & 2.553 & 41.594 & -4.1 & 35.2\tabularnewline
IPSW & 3.504 & 12.013 & 156.306 & -33.7 & 77.2 & -1.697 & 12.673 & 163.171 & -29.3 & 87.0\tabularnewline
AIPSW & -1.240 & 3.369 & 12.866 & -27.8 & 81.8 & -2.060 & 4.721 & 26.490 & -37.8 & 77.2\tabularnewline
AIPSW(S) & -0.084 & 0.676 & 0.463 & -22.7 & 87.6 & -0.023 & 0.742 & 0.550 & -31.9 & 85.6\tabularnewline
SBW & 0.693 & 8.244 & 68.304 & 4.3 & 93.6 & -1.361 & 9.863 & 98.928 & -3.1 & 92.6\tabularnewline
CW & 0.894 & 11.174 & 125.400 & -18.2 & 84.2 & -1.375 & 12.375 & 154.728 & -17.2 & 85.6\tabularnewline
ACW-t & -0.146 & 3.505 & 12.283 & -14.6 & 91.2 & -1.298 & 4.190 & 19.205 & -20.5 & 87.2\tabularnewline
ACW-t(S) & -0.049 & 2.343 & 5.479 & -33.9 & 95.0 & 0.042 & 1.119 & 1.252 & 9.9 & 94.8\tabularnewline
ACW-t$\left(\mbox{S}^{O}\right)$ & -0.055 & 0.694 & 0.483 & -1.6 & 95.0 & 0.001 & 0.675 & 0.455 & -10.7 & 93.4\tabularnewline
ACW-b & 0.160 & 3.318 & 11.010 & -15.1 & 92.2 & -1.261 & 3.792 & 15.939 & -22.0 & 87.2\tabularnewline
ACW-b(S) & 0.133 & 2.169 & 4.715 & 0.2 & 97.2 & 0.091 & 1.141 & 1.307 & 63.8 & 97.2\tabularnewline
ACW-b$\left(\mbox{S}^{O}\right)$ & -0.032 & 0.574 & 0.329 & -5.3 & 94.6 & 0.034 & 0.565 & 0.320 & -3.5 & 95.2\tabularnewline
\bottomrule 
\end{tabular}
}
\end{table}

\begin{figure}[h]
\centerline{
\includegraphics[width=6in]{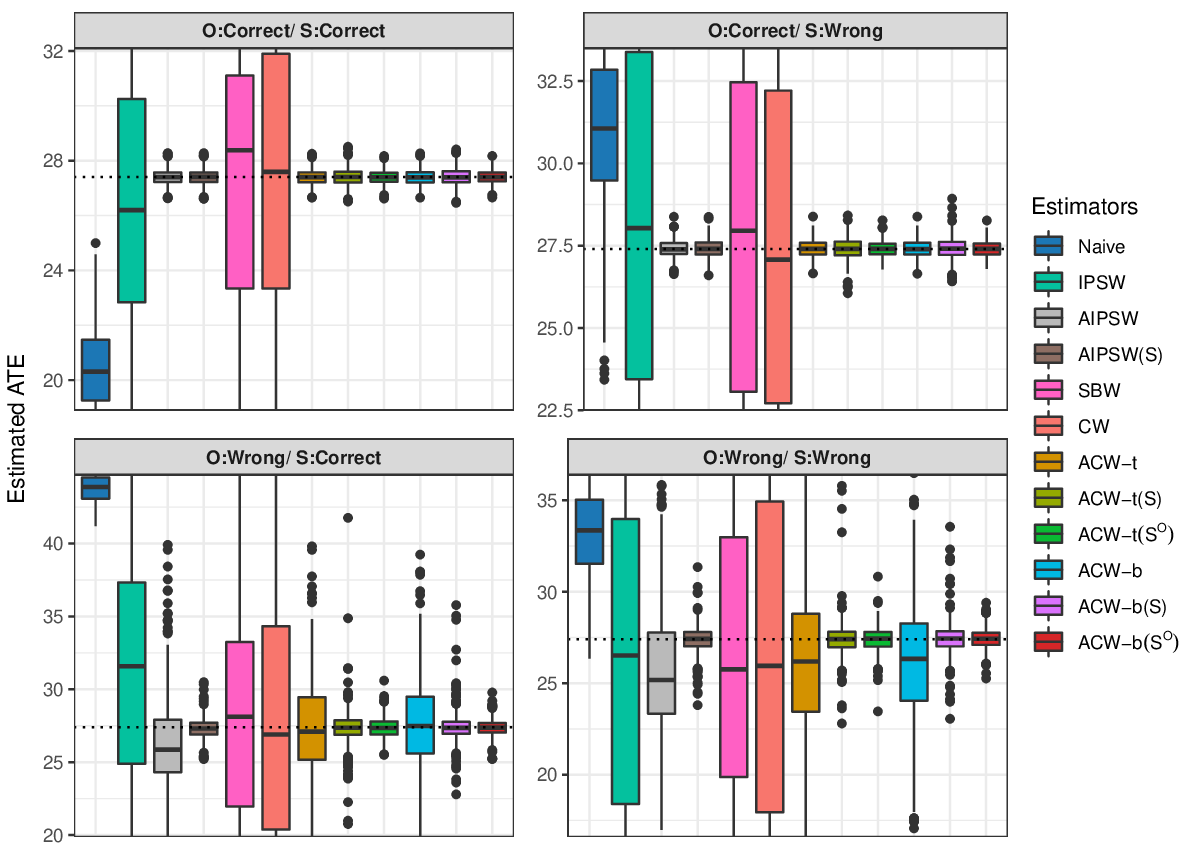}}
\caption{Boxplot of estimators with increased observational study sample size. We removed a few outliers for visualization. \label{fig:comparelarge}}
\end{figure}

\end{document}